\documentclass[twocolumn,twocolappendix]{openjournal}

\pdfoutput=1 

\usepackage{amsmath,amstext}
\usepackage[T1]{fontenc}
\usepackage{apjfonts}
\usepackage{ae,aecompl}
\usepackage[utf8]{inputenc}
\usepackage{hyperref}
\usepackage[figure,figure*]{hypcap}

\usepackage{booktabs}
 
\setcitestyle{square}

\usepackage{lineno}

\newcommand{\Msol}{\ensuremath{\mathrm{M}_{\odot}}}

\shorttitle{Unknown Unknowns}
\shortauthors{Peter Hatfield}

\begin{document}
\title{Quantification of Unknown Unknowns in Astronomy and Physics}
\author{Peter Hatfield}
\email{astropeterhatfield@gmail.com}
\affiliation{Astrophysics, University of Oxford, Denys Wilkinson Building, Keble Road, Oxford, OX1 3RH, UK}

\begin{abstract}
Uncertainty quantification is a key part of astronomy and physics; scientific researchers attempt to model both statistical and systematic uncertainties in their data as best as possible, often using a Bayesian framework. Decisions might then be made on the resulting uncertainty quantification - perhaps whether or not to believe in a certain theory, or whether to take certain actions. However it is well known that most statistical claims should be taken contextually; even if certain models are excluded at a very high degree of confidence, researchers are typically aware there may be systematics that were not accounted for, and thus typically will require confirmation from multiple independent sources before any novel results are truly accepted. In this paper we compare two methods in the astronomical literature that seek to attempt to quantify these `unknown unknowns' - in particular attempting to produce realistic thick tails in the posterior of parameter estimation problems, that account for the possible existence of very large unknown effects. We test these methods on a series of case studies, and discuss how robust these methods would be in the presence of malicious interference with the scientific data.
\end{abstract}



\section{Introduction}

Uncertainty quantification is a key part of the scientific method and modern physics and astronomy (\cite{Andrae2010,Barlow2002}). When data is taken from an experiment or observation, as well as finding what model is most supported by the data (be that an estimation of a numerical value, for example measuring Hubble's constant, or deciding between two physical models for example does the data favour Newtonian gravity or General Relativity), we also typically want an estimation of how \textit{confident} we are in that inference. Were the measurements very tightly constraining, or are we still very uncertain between different possibilities? Uncertainty quantification is a large field, but one common framework popular in physics is \textit{Bayesian statistics} (\cite{VonToussaint2011}). For parameter estimation a prior is chosen over a set of parameters, a model of how the data is produced is used to calculate the liklihood, and they are combined to produce a posterior e.g. \cite{Li2016}. The resulting posterior is however very highly dependent on the choice of the model; if the model is false (misspecified) then the resulting posterior is likely to be misleading (\cite{Draper1995,Walker2013}).

\smallskip

Posteriors produced in this way are thus highly conditional on the model used being true, and essentially capture uncertainty from the `known unknowns'. `Known unknowns' are any sources of uncertainty the researcher is aware of and includes in the analysis - they might be statistical, systematic, numerical, or any of the many other sources of uncertainty. For actually making decisions however this may not be sufficient (\cite{Draper1987,Grabo2002}). Real data can often have missing parts, be censored in some way and be correlated in ways we had not anticipated (\cite{Osborne2012}). Making a decision might mean a number of things; in the context of blue sky scientific research it might correspond to deciding whether or not to believe in a new particle, for applied applications it might be a choice of whether or not to launch a crewed spaceflight. For either scenario we wish to know the \textit{true} uncertainty\footnote{By which is meant uncertainties that are well calibrated in the sense that things assigned $x$\% chance of happening happen $x$\% of the time, but also in the sense that one would be prepared to truly rely on the calculation - would you be prepared to bet your goldfish, your dog, or your life on the estimate? \url{https://blog.sciencemuseum.org.uk/the-science-of-the-multiverse/}}, and remove the conditionality on the truth of the model. We would wish to quantify the possibility of the `unknown unknowns' \textit{as well} as the known unknowns\footnote{`...as we know, there are known knowns; there are things we know we know. We also know there are known unknowns; that is to say we know there are some things we do not know. But there are also unknown unknowns - the ones we don't know we don't know.' - United States Secretary of Defense Donald Rumsfeld, 12th February 2002}. Quantifying unknown unknowns is a challenging task, and is to some degree an ill-defined problem (\cite{Taleb2012}) - they might be statistical or systematic (\cite{Sinervo2003}) - here we define an unknown unknown as anything not included in the modelling or otherwise anticipated. However one possible approach to at least make some constraints on the possible sizes of unknown unknowns is to use \textit{multiple independent probes} of the physics; if multiple methods give consistent answers then we are much more likely to believe the results\footnote{C.f. the Swiss Cheese Model in risk anaysis}, as we would expect systematics to be uncorrelated on the different data sets. Conversely if multiple independent probes give different answers/predictions then we should conclude that the physics is poorly understood and thus should have large uncertainty about conclusions, and expect there to be a reasonable chance of currently unanticipated processes impacting outcomes. 

\smallskip

One approach in the literature to formalise inference from multiple possibly inconsistent probes is to introduce a hierarchical model where hyperparameters describe in some way `the probability of being wrong' (in this paper we focus on \cite{Press1996} and \cite{Bernal2018})\footnote{Although there are a few other approaches, \cite{Rajput1992,MacMahon2004,Pritychenko2012,Pritychenko2017}}. This is a `model expansion' approach (\cite{Kennedy2001}). These parameters can then be inferred at the same time as the parameters of interest, and then marginalised out as nuisance parameters. This approach is essentially an evolution of the well known methods for treatment of outliers/erratic data\footnote{See also the concept of robust statistical methods}, where the noise on measurements is unknown, or allows individual data points in curve fitting are assigned probabilities of being an outlier (\cite{Sivia2006}) - it is simply applying these methods at the level of whole probes rather than individual data points. Accounting for the possibility of unknown systematics can to some degree be thought of as a extension of ``Cromwell's rule''\footnote{Named by  Dennis Lindley (referring to Oliver Cromwell, who wrote to the General Assembly of the Church of Scotland on 3 August 1650 `I beseech you, in the bowels of Christ, think it possible that you may be mistaken', \cite{Jackman2009})} that prior beliefs should always be non-zero (even if very small) - we should also always allow a non-zero probability that our measurements are flawed in some currently unknown manner.

\smallskip

Whether or not attempting to account for unknown unknowns is useful or necessary depends on the context, for example it depends how costly an extreme deviation from predictions is; if the occasional bad prediction is unproblematic then allowing the extra freedom for unknown systematics may be unnecessary. In addition the model-dependent estimates and uncertainties are still the primary calculations for the purposes of advancing science, as comparing different estimates is part of the process of improving our understanding of the physics and converting unknown unknowns to known unknowns.  However for situations where decisions must be made and tail risks are problematic we should attempt to include a possibility that the scientific models used are flawed. Possible physics cases where we suggest accounting for tail risk from unknown physics could be important include crewed spaceflight (\cite{Shayler2000}), nuclear weapon anti-proliferation and treaty verification (\cite{Niemeyer2020}), climate change and energy security (\cite{Katzfuss2017}), Earth Observation for disaster monitoring (\cite{LeCozannet2020}), space weather (\cite{Camporeale2019a}) and space debris collisional cascades (\cite{Walker1998}).

\smallskip

In this paper in section \ref{sec:methods} I briefly describe the methods of \cite{Press1996} and \cite{Bernal2018}, in section \ref{sec:case_studies} I apply the methods to three case studies, in section \ref{sec:malicious} I discuss possible application of the methods to protect against malicious interference of the data, in section \ref{sec:discussion_various} I discuss practical aspects of the methods, and finally in section \ref{sec:conclusion} I conclude. In the Appendix I describe a simple real world example, applying the methods to modelling the evolution of the COVID-19 pandemic.

\section{Methods} \label{sec:methods}

We highlight two methods that seek to model the possible existence of unknown systematics. Both were developed in the context of cosmological parameter estimation, and were written in the context of emerging tensions (\cite{DiValentino2021}) between estimates of the Hubble constant (albeit at the 20\% level in the 1990's, and at the few percent level in the 2010's/2020's) from different probes of the physics (both independent telescopes, as well as different physical methods). Both meta-analysis approaches rely on essentially introducing new parameters that describe how the probes might be biased, that are then constrained alongside the parameters of interest, and then marginalised out to give final cosmological parameter posteriors. Hierarchical Bayesian Models have been used in many areas of physics e.g. \cite{Feeney2018,Hall2021, Hinton2019,Huang2010,Nayak2020}.

\smallskip

\cite{Hobson2002} note that introducing parameters that describe unknown systematics is introducing new parameters, so is effectively increasing the model complexity, and therefore are not necessarily justified, and use calculations of Bayes Factors to justify whether the new parameters are necessary or not. However in this text we mainly consider scenarios where we have good reason to believe that there are unknown systematics, so in this text we will largely assume the extra model complexity is justified (many authors have considered different ways to test if data is internally consistent e.g. \cite{Kohlinger2019}). However an approach that accounts for this may be needed to get realistic uncertainties for intermediate cases where there could be moderate unknown systematics, but the measurements are not in terrible disagreement.

\smallskip

We leave the definition of `independent probe' slightly ill-defined throughout this text (\cite{Bernal2018} refer to independent classes of data). Ideally probes would be completely different explorations of the phenomena; for example in the cosmological example, the Cosmic Microwave Background and Supernovae redshift measurements are both highly independent ways of studying the physics of the expansion of the Universe.  Treating two different telescopes both measuring the Cosmic Microwave Background as separate probes might help identify instrument systematics, but not unidentified systematics in our understanding of the physics. Many other fields have identified the need for highly independent qualitatively different sources of data - the notion of an `independent probe' could be viewed as being similar to the doctrine of `dual phenomenology', \cite{Futter2018}, for attack warning systems.

\smallskip

For illustrative purposes Figure \ref{fig:toy_figure_1} shows the posteriors from the two methods for a simple example case of two probes, with means $\mu = \pm 1$ and standard deviations $\sigma=0.5$. Scripts similar to those used for this toy model are available at \url{https://github.com/pwhatfield/unknown_unknowns/}.

\subsection{Non-Informative Prior}

All choices of prior involve some judgement; sometimes there might be well physically motivated choices of prior that can be used, or a prior can be used based on previous experimental findings. However there are also well established methods to assign sensible non-informative priors when you are highly ignorant e.g. the Jeffreys prior, \cite{JEFFREYS1946}. For our purposes this normally amounts to selecting a uniform prior for a location-like parameter and a uniform prior in log-space for a scalar positive parameter. For a a Bernoulli distribution we use the appropriate beta distribution ($\beta(\frac{1}{2},\frac{1}{2})=[x(1-x)]^{-\frac{1}{2}}$) for the prior (e.g. \cite{Gelman2014}), truncated to lie between 0.05 and 0.95 (a semi-arbitrary choice that depends on the problem in question, to always allow some probability of a probe being `right' or `wrong').

\subsection{Conventional Application of Bayes' Theorem}

Ordinarily if there are multiple independent probes of a physical phenomenon, one can use Bayesian updating, the experimentalist/observer starts with some prior, sees the data from Probe 1, constructs a likelihood, finds their posterior, which becomes the new prior and continues iterating. The order in which the data are seen doesn't matter, so the likelihood from the data as a whole can be considered to be the likelihoods from each individual probe multiplied together. We will subsequently refer to this as Conventional Bayes' (CB).

\subsection{Press 1996}

\cite{Press1996} (subsequently P96) introduces freedom to the modelling for each probe to be either `right' or `wrong'. It introduces a parameter\footnote{Not given a symbol in P96} $\phi$ which describes the probability of a given measurement of a parameter being `good' or `bad'. Marginalisation is done over $\phi$, and the vector $V$ of $N$ truth values representing the probabilities of different measurements being right or wrong. So if the original model had $K$ parameters to apply Bayes' theorem to, the new model to use Bayes' theorem over is $K+1$ parameters and $N$ binary values. $\phi$ itself requires a prior, and may be well known (e.g. we know in advance that measurements are only `right' about 25\% of the time, or it may be poorly known in which case we might allow it to take a broad range of values)\footnote{It should be emphasised that in this method, and in none of the case studies considered in section \ref{sec:case_studies} in no way is is a low value of $\phi$ indicative of poor scientific work; more it is indicative of a challenging area of physics with lots of `surprises'. Similarly `good/bad' and `right/wrong' are not value judgements.}. A strong prior of $\phi=1$ would correspond to reverting to CB. A strong prior of $\phi=0$ would correspond to never trusting any data and sticking resolutely to prior beliefs. $V$ similarly requires a prior, which has a natural one based on the prior on $\phi$, but in principle could be chosen in a bespoke manner (say some individual probes were known to be secure, but others were not). $\phi$ and $V$ can be marginalised out to get a $K$-dimensional posterior over the parameters of interest. Similarly the parameters of interest can be marginalised out to get posteriors over $\phi$ (and $V$). P96 illustrates clearly that this method prevents individual measurements with unrealistically small uncertainties dominating other measurements that are in greater agreement with each other - and gave a posterior for the cosmological parameter $H_0$ that many years later has been shown to be accurate. We make the minor modification to P96 of reverting to the parameter prior rather than a very wide Gaussian of width $S$ when the measurement is considered `bad' but this change doesn't have a huge impact on results.

\smallskip

It can be seen in Figure \ref{fig:toy_figure_1} that P96 posteriors are essentially mixture models of the $2^N$ (where $N$ is the number of probes) probability distributions that correspond to each possible set of `right' and `wrong' assignments, weighted appropriately (c.f. Bayesian model averaging).

\subsection{Bernal and Peacock 2018}

\begin{figure}
\includegraphics[scale=0.6]{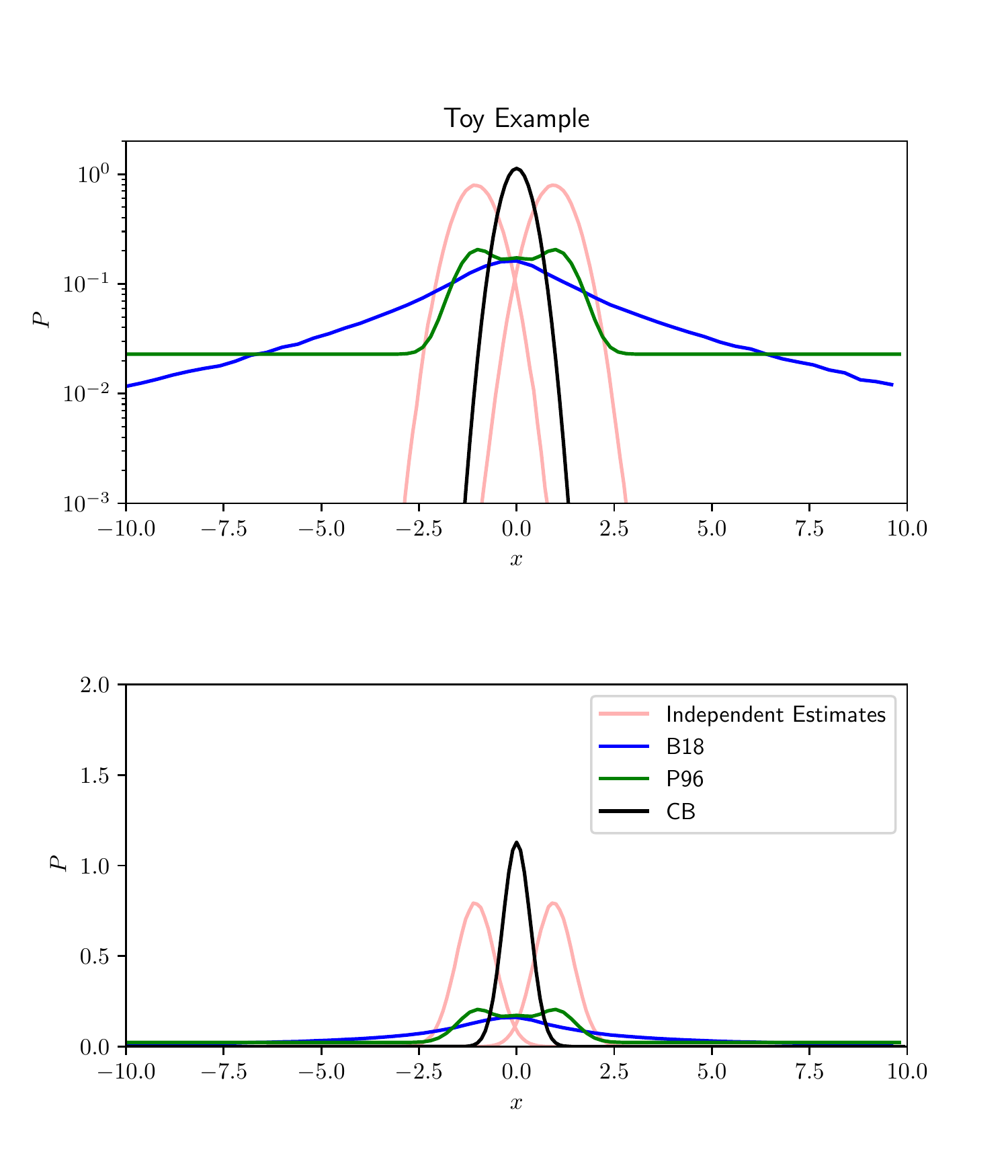}
\caption{The posteriors for a toy example of two probes with means $\mu=\pm1$ and standard deviation $\sigma=0.5$. Black shows conventional posteriors, green shows the posteriors from P96, blue shows B18 posteriors and the light red are the distributions from the individual separate probes. The top and bottom sub-plots show the same distributions, with a log-scale for the upper plot, and a linear scale for the lower plot.}
\label{fig:toy_figure_1}
\end{figure}

\cite{Bernal2018} (subsequently B18) assumes the calculated likelihood from each probe is distorted both in location and dispersion. Each likelihood is modelled to be shifted in parameter space, and rescaled, by some unknown quantities. Location shifts are assumed to be sampled from a Gaussian with mean zero, and unknown standard deviation (that becomes a new unknown parameter), and rescalings are sampled from an exponential distribution. The values of the shifts, the standard deviation of the distribution they are sampled from, and the rescalings are new unknowns and are given uninformative priors, and are then included in the application of Bayes' theorem, and then subsequently marginalised out. A similar process is applied to the dispersion of the likelihoods. For the estimation of 1D parameters it therefore introduces $2N+1$ new parameters, $N$ for the location, one for the standard deviation of the distribution from which shifts are sampled, and $N$ for dispersion. B18 also describes in detail how to deal with inference in higher dimensional spaces and how to put priors on covariance of parameters, neither of which we consider here. They apply this to a number of cosmological data sets, and find that more realistic uncertainties are found, and in particular illustrate that systematics on both location and dispersion are required. B18 posteriors reduce to the Student-$t$ distribution\footnote{See \cite{Zholud2014} for a discussion of appropriate tails to expect in different small sample scenarios.} in the limit of multiple inconsistent measurements with unreasonably small uncertainties, see Figure \ref{fig:toy_figure_1}.

\smallskip

We note that B18 does not permit inferences when there is only one probe, as then it is very difficult to constrain the unknown Gaussian parameters with only one measurement, so if only one probe is available it would be necessary to use user-defined priors on those parameters. B18 also perform detailed working checking for convergence of MCMC chains, which we do not discuss in depth here but is important for proper sampling of tails. We largely use the pre-set approach of the BACCUS implementation made available by the authors of B18\footnote{\url{https://github.com/jl-bernal/BACCUS}}.

\subsection{Comparison}

The different resulting posteriors from P96 and B18 can be seen in Figure \ref{fig:toy_figure_1}. Firstly both give wider central estimates, and fatter tails, than conventional Bayes' Theorem. It can also be seen that for intermediate deviations ($\pm3$ in the toy example) B18 gives the greater probability, as the possibility of shifts on the estimates is permitted. However for larger deviations P96 gives greater tail probabilities, as B18 finds large shifts to be unlikely, but P96 accepts the possibility of both probes being completely `bad'. In addition P96 and B18 both also capture the benefit of having additional low-precision probes - adding new probes with large uncertainties don't change the central estimates much, but \textit{does} reduce the probability in the tails.

\smallskip

The hierarchical models described in P96 and B18 are themselves models, and thus subject to systematics - they will not perfectly represent the way that unknown unknowns are generated. Thus even these methods likely still won't capture the true uncertainty of tail probabilities - for example if the systematics on the uncertainties on individual probes are correlated unbeknown to us. In particular they also disagree, illustrating that they aren't perfect representations of the true uncertainty. Note in particular that the nature of the unknowns is presumed to be different in the two methods. P96 assumes that if a probe is not `correct' then it is \textit{completely} wrong and essentially reverts to the prior - colloquially it allows for the possibility that measurements may be \textit{qualitatively} wrong. B18 conversely assumes that no method is completely wrong, merely that the methods and uncertainty calculations can be very badly biased - colloquially assuming that measurements may be \textit{quantitively} wrong. Thus it can be seen that the main factor in estimating tail risks is whether or not one believes qualitative or quantitative risks are dominant.

\section{Case Studies}  \label{sec:case_studies}

\begin{figure*}
\includegraphics[scale=0.6]{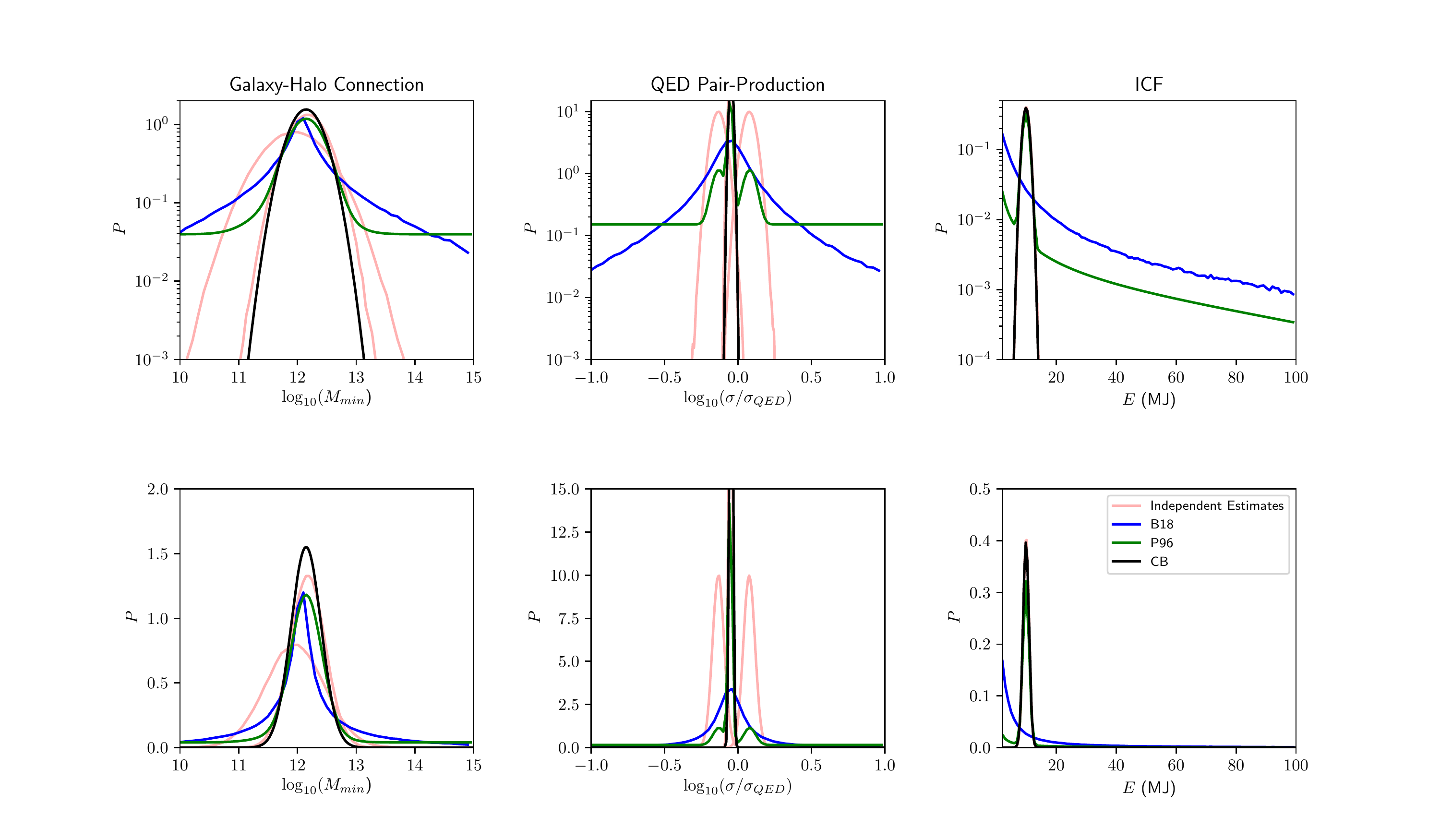}
\caption{The posteriors for the three physics case studies, galaxy-halo connection (left two sub-figures), pair-production (central two sub-figures) and ICF (right two sub-figures). Probabilities are shown in both log (1st row) and linear space (2nd row). Black shows conventional posteriors, green shows the posteriors from P96, blue shows B18 posteriors and the light red are the distributions from the individual separate probes.}
\label{fig:combined_figure_1}
\end{figure*}

\begin{figure*}
\includegraphics[scale=0.6]{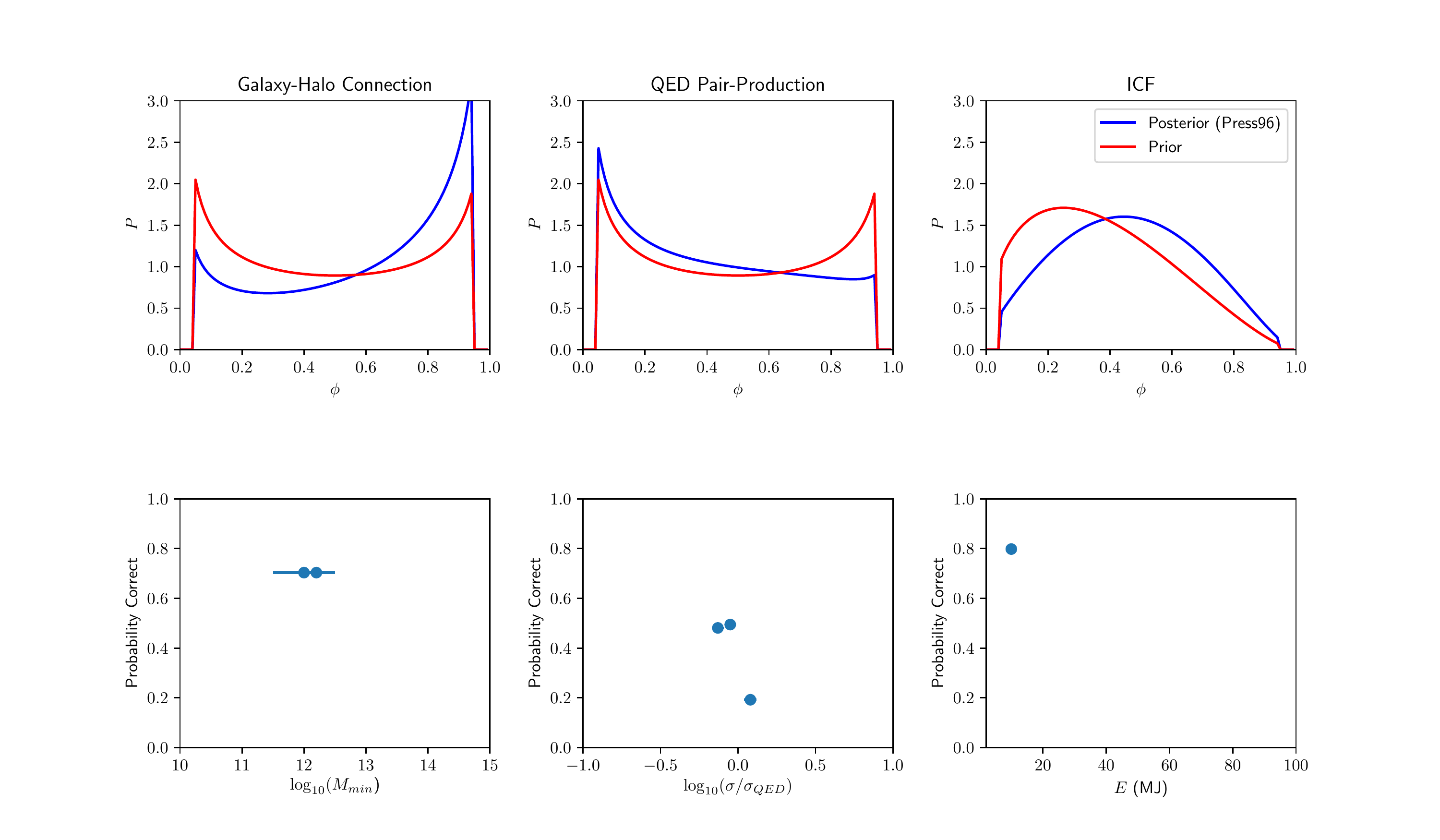}
\caption{The parameters describing the probabilities of measurements being correct in the P96 model for for the three physics case studies, the galaxy-halo connection (left two sub-figures), pair-production (central two sub-figures), and ICF (right two sub-figures). The 1st row shows the global probabilities (the red curves corresponding to the prior, and the blue curves to the posterior), the 2nd row shows the probabilities for individual probes.}
\label{fig:combined_figure_2}
\end{figure*}

In this section we apply the methods of P96 and B18 to three scientific cases studies where there is the potential for unknown physics. In all three cases the goal is to illustrate the methods, rather than to make specific claims in each of the domains; researchers may wish to do their own versions of the analysis with their own likelihoods and priors. In the Appendix we also discuss an example relevant to the ongoing problem of modelling the COVID-19 crisis. Figure \ref{fig:combined_figure_1} shows the posteriors acquired from conventional application of Bayes' Theorem, P96 and B18. Figure \ref{fig:combined_figure_2} shows the implied global probabilities of being `wrong' and the probabilities for individual probes from P96. Table \ref{tab:percentiles} shows the medians, the 16th and 84th percentiles (`$1-\sigma$') and 5th-95th percentile range. In all case studies we use Gaussian likelihoods for each individual probe, however the methods can be applied regardless of the form of the liklihood.

\smallskip

We comment on the specific cases studies in the follow sub-sections, but here briefly summarise some common features. Firstly the central estimate of uncertainty is always increased, normally by a factor of about 2. Secondly P96 and B18 give increased estimates of tail probabilities by many orders of magnitude. Thirdly P96 and B18 don't normally perfectly agree, but normally only differ by about a factor of ten, which is comforting given the challenges of extrapolating to low probabilities.

\begin{table*}

\caption {Central estimates (median and 16th-8th percentile range) and 90\% Credible Interval (5th-95th percentile range) for our three case studies and 4 probability distributions, the original prior, Conventional Bayes (CB), P96 and B18.}
\begin{center}
\begin{tabular}{@{}ccccc@{}}
\toprule
	& 	& \textbf{Probability Distribution}	&  \\
\midrule
& \textit{Prior}& \textit{CB}& \textit{P96}& \textit{B18}\\
\midrule
\textit{Galaxy-Halo Connection} ($\log_{10}[M_{\mathrm{min}}/\Msol]$)\\\midrule
Central Estimate & {$12.5^{+2}_{-2}$} & {$12.2^{+0.3}_{-0.3}$} & {$12.3^{+0.4}_{-0.4}$} & {$12.2^{+0.7}_{-0.5}$}\\
90\% CI & {[10.3,14.8]} & {[11.8,12.6]} & {[11.2,13.8]} & {[10.9,13.8]}\\
\midrule
\textit{Matter from Light} ($\log_{10}[\sigma/\sigma_{\mathrm{QED}}]$)\\\midrule
Central Estimate & {$0.0^{+0.7}_{-0.7}$} & {$-0.03^{+0.01}_{-0.01}$} & {$-0.02^{+0.2}_{-0.1}$} & {$0.00^{+0.2}_{-0.2}$}\\
90\% CI & {[-0.9,0.9]} & {[-0.05,-0.01]} & {[-0.64,0.68]} & {[-0.36,0.4]}\\
\midrule
\textit{Inertial Confinement Fusion} (MJ)\\\midrule
Central Estimate & {$14^{+40}_{-10}$} & {$11^{+1}_{-1}$} & {$12^{+1}_{-2}$} & {$9^{23}_{-4}$}\\
90\% CI & {[2.4,82]} & {[9,12]} & {[7,33]} & {[4,73]}\\
\bottomrule
\end{tabular}\label{tab:percentiles}
\end{center}
\end{table*}

\subsection{Galaxy-Halo Connection}

P96 and B18 were both developed for the problem of estimating cosmological parameters. Constraining the galaxy-halo connection is a closely related problem in astrophysics, where the relationship between galaxies and dark matter haloes\footnote{`Clumps' of dark matter, within which galaxies are believed to form and be found.} is measured (for example what is the mass of a galaxy as a function of its dark matter halo mass). This is particularly challenging as dark matter is believed to be largely invisible, so instead a large number of indirect methods are used. There are several such methods to probe the relationship, but each has several differing systematics, some of which are known, some of which are likely unknown.

\smallskip

Here we apply P96 and B18 to modelling from \cite{Hatfield2016}, which used data from the VIDEO survey (\cite{Jarvis2013}) to constrain the galaxy-halo connection. \cite{Hatfield2016} used Bayes' Theorem and MCMC to constrain the parameters of a 5-parameter model using two probes, galaxy number counts, and galaxy clustering, by simply multiplying the two likelihoods together (Conventional Bayes'). Here we seek a posterior that accounts for unknown systematics, by fitting the same parameters to first just the number counts, and then second to just the clustering measurements. Then we combine the resulting posteriors using P96 and B18. We only display one of the parameters, $M_{\mathrm{min}}$, from the modelling, although the methods can be extended to the multi-dimensional distributions. We use a prior uniform in log-space over $M_{\mathrm{min}}$ as mass is a scalar parameter.

\smallskip

The new posteriors in figure \ref{fig:combined_figure_1}  is not too dissimilar to the original posterior, except it has thicker tails, and has a larger central uncertainty (16th and 84th percentiles). This method is likely sensitive to systematics that are particular to each probe (for example a non-uniform galaxy detection efficiency across the sky\footnote{Specifically such an effect beyond that already accounted for.} would impact clustering substantially but not the number counts) but not those common to all probes e.g. redshift systematics (\cite{Hatfield2019}). Figure \ref{fig:combined_figure_2} shows after studying the two probes we ought to favour slightly higher $\phi$ values (as the probes were consistent).

\smallskip

In reality we have other probes of the galaxy-halo connection; for example \cite{Coupon2015} also add galaxy-galaxy lensing to the modelling, so in practice we have good reason to believe that the tails are thinner than portrayed here. However when a similar analysis is done in future on all-sky surveys such as will be performed at the Rubin Observatory (\cite{Ivezic2019}), there will be $\sim$10,000 times as much data, so the nominal uncertainties will be much smaller and real tensions are likely to emerge. Current studies already disagree with each other to a greater degree than quoted error bars, so there are already some constraints on the sizes of unknown systematics (\cite{Ishikawa2020})\footnote{Again, researchers are largely aware of this; the values quoted in the literature are more like conditional probabilities, conditional on a large number of assumptions being true. P96 and B18 seek to give more realistic probability distributions, that one might use if having to make a decision on believing a certain model or not.}. To some degree this will be mitigated by improved modelling of non-linear galaxy clustering, and separating the galaxies into smaller bins based on their properties. However constraints are in general going to be much tighter, so it will likely make sense to combine probes in ways that account for unknown systematics, particularly as we likely don't anticipate being able to model the complex baryonic physics of galaxy formation to the precision of the surveys by the time the surveys are complete. Finally we would emphasise the possibility of radical uncertainty even beyond what P96 and B18 allow for, in this case it is unlikely, but possible, that dark matter doesn't exist - in which case estimates of halo masses are meaningless.

\subsection{Matter from Light}

When matter meets anti-matter, the particles annihilate, producing gamma rays ($E=mc^2$, `light from matter'). Less well commonly known is that it is theoretically predicted that two photons are able to `collide' and produce an electron and positron (\cite{Breit1934}, $m=E/c^{2}$, `matter from light', the Breit-Wheeler process). This phenomenon can be probed in several ways in the laboratory, and is also believed to happen in several astrophysical scenarios. Several experiments and astrophysical observations have attempted to measure the Breit-Wheeler  process, and in some of the probes Beyond Standard Model (BSM) physics has been suggested. If true this would be a highly impactful result, indicating a fundamental change in our understanding of quantum electrodynamics (QED). However these methods all have large numbers of unknown systematics, and probe the physics in different ways, making it important to combine the probes in a robust way, to confirm that a detection of novel physics is more likely than the presence of an unknown systematic (a point made in B18). Historically many $5\sigma$ claims of non-standard model physics have latter found to be inaccurate not because they were one-in-a-million fluctuations, but because of some previously unappreciated physics e.g. the size/impact of cosmic dust in the case of possible gravitational wave detection by BICEP-2 (\cite{Cowen2015}) and problems in a fiber optic cable and clock oscillator in the case of possible superluminal measurements from OPERA (\cite{Adam2012}, c.f. \cite{Reece2009}).

\smallskip

Here we consider three experiments/observations that have attempted to measure the possible existence of pair-production, and their consistency with QED; a Stanford Linear Accelerator Center (SLAC) experiment (non-linear pair production, $\log_{10}(\sigma/\sigma_{\mathrm{QED}})=-0.13\pm0.04$) \cite{Burke1997}, experiments at the Large Hadron Collider (LHC, linear, but of virtual-virtual photon pairs, $\log_{10}(\sigma/\sigma_{\mathrm{QED}})=0.08\pm0.04$), \cite{Abbas2013}, and through observation of high energy gamma rays (real photons, but subject to unknown astrophysical systematics, $\log_{10}(\sigma/\sigma_{\mathrm{QED}})=-0.05\pm0.012$), \cite{Horns2012,Biteau2015}. These represent highly independent probes of the underlying physics and would not expect any correlation in unknown systematics. We see that the robust posteriors are perfectly consistent with the Standard Model (Figure \ref{fig:combined_figure_1} and Table \ref{tab:percentiles}), suggesting there is no reason yet to believe there is anything fundamentally wrong with QED. However the uncertainty is still up to an order of magnitude, so modifications to the theory of this size are still plausible. In addition it can be seen in Figure \ref{fig:combined_figure_2} that we would now favour slightly lower values of $\phi$, due to the level of inconsistency between the probes.  Further experiments in photon-photon physics for example 2018 experiments (\cite{Kettle2021}) at the Gemini laser (based on \cite{Pike2014}) will hopefully help reduce the size of the tails. Finally many science challenges require the combination of laboratory and observational data (e.g. \cite{Evans2019}) - we have shown here that Bayesian model expansion methods are one robust way to do this.

\smallskip

This case study illustrates the difference between qualitative (captured by P96) and quantitative (captured by B18) systematics on the probes. Quantitative unknown systematics might take the form of the energy of the photons involved being different than calculated, or perhaps the efficiency of the detector used was different than modelled. Qualitative unknown systematics might correspond to a scenario where mistakenly the detected electrons and positrons were not actually produced by the Breit-Wheeler process - perhaps the experiment is mistakenly detecting particles produced by the Bethe-Heitler or the Trident process, in which case the number of particles detected would have no connection to the Breit-Wheeler cross-section.

\smallskip

The physical meaning of the `cross section' as implemented here is debatable, as all the experiments probe pair-production at slightly different energies and forms of the process i.e. one could conceive of BSM theories that modified the QED cross-sections to a much greater degree at some energies but not at others. There is thus a trade off - the more highly independent the probes are, the greater the risk that they are no longer truly comparable. However analysis like this is likely necessary if we want to prove the existence of non-Standard Model physics (e.g. axions) via experiments subject to large unknown systematics (like high-powered lasers). Demonstrating a non-Standard Model measurement of the cross-section from just one of these methods would likely not be believable; if several highly independent probes started to indicate a deviation from the SM, and an uncertainty quantification that incorporated unknown systematics still supported deviation from SM, then it may start to gain credibility\footnote{See also medical meta-analyses, \cite{Haidich2010}.}. It should also be noted that unlike the Galaxy-Halo example, some of the parameter values have greater importance than others e.g. $\log_{10}(\sigma/\sigma_{\mathrm{QED}})=0$ corresponds to QED being correct. Thus a larger analysis might involves calculating Bayes' Factors between QED and a free parameter model that included unknown unknowns.

\subsection{Inertial Confinement Fusion}

Inertial Confinement Fusion (ICF) is one possible pathway to nuclear fusion as an industrial power source - as well as the only way to test a range of astrophysical and cosmological physics in the lab, \cite{Rose2021}. The main ICF facility in the world is the National Ignition Facility (NIF) at Lawrence Livermore National Laboratory (LLNL) in California, USA \cite{Lindl2004}. NIF is a $\sim2$MJ laser, and before being built was expected to reach ignition (in the context of energy production, achieving the goal of more energy given out than put in), but unfortunately this target was not reached (\cite{Hurricane2016a}), although a huge number of engineering and scientific accomplishments have been made. Upgrading NIF to a larger system is now under consideration (\cite{Rose2020,Rose2021}), with current work attempting to quantify a) how many MJ a successor laser would have to be to reach ignition and b) what the uncertainty on that prediction is.

\smallskip

There has recently been increased interest in the use of statistical methods for ICF (\cite{Hatfield2021}). It was emphasised in \cite{Kasim2019} that using Bayes' Theorem was necessary to avoid underestimating uncertainty for the field of high energy density physics (of which ICF is a part). \cite{Gaffney2019} pioneered the use of Bayesian calibration taking into account data from multiple NIF shots to make predictions on energy yield from future shots (see also \cite{Ruby2021}). Finally \cite{Osthus2019} illustrated the combination of multiple different possible extrapolation models to get more realistic uncertainties on predictions. But what is the probability of none of the models being correct?

\smallskip

As discussed, uncertainties typically are conditional on the assumptions on the model being correct (as \cite{Nakhleh2020} emphasise). The history of ICF unfortunately has been subject to large numbers of unanticipated physics effects each time a new scale is reached, so we likely should be wary of such possibilities in future\footnote{As noted in a lecture by Professor Jeremy Chittenden, `Thermonuclear fusion versus Murphy's Law', \url{https://www.youtube.com/watch?v=XRtMayvnLoI}.}.  In the context of P96 we might expect the value of $\phi$ for the field to be quite low. For NIF ~1MJ was predicted to be the ignition scale (\cite{Hammer2005}), but `unknown unknowns' including the impact of greater than expected 3D imperfections and mixing effects prevented it from reaching its goal at 2MJ. Previous lasers\footnote{\tiny{\url{https://lasers.llnl.gov/10-years-of-dedication/laser-leadership}}} have had similar challenges; for example SHIVA ($\sim$0.01MJ) and NOVA ($\sim$0.2 MJ, which was expected to achieve gain before it was built), were impacted by a range of unanticipated phenomena/known phenomena that were much more problematic than anticipated, e.g. variation in laser intensity in the beams, filamentation, laser coupling with hot electrons, Non-Local Thermodynamic Equilibrium (NLTE) atomic physics for the high-Z hohlraum walls, and many aspects of target performance (\cite{Lindl1995}). Estimates of ignition scale often increased by many factors in just a few short years as more was understood about the problem\footnote{Some have argued there is a similar problem with the age of the Universe: \url{https://www.lockhaven.edu/~dsimanek/cutting/ageuniv.htm}}. It is beyond the scope of this paper to consider further, but it has been noted in the literature that estimates with uncertainties should not change as fast at this if you are betting on them (i.e. if odds have changed so much to date, you should also expect them to change in future) - interesting future work could investigate the statistical properties of the time series of ICF ignition scale forecasts (e.g. the Martingale condition, \cite{Taleb2018}). Current estimates of ignition scale are at approximately 5-10MJ (depending on models used); \cite{Randewich2020} note that it is `unlikely to be above 5 MJ\footnote{The situation is slightly complicated by various possible definitions of what a facility reaching ignition entails.} unless unknown physics kicks in' - hopefully this case study can illustrate how to calculate the rough probability of unknown physics becoming relevant.

\smallskip

With regards to choice of prior on ignition scale, prediction in ICF is very challenging, but has one major advantage (as compared to other pathways to fusion like magnetic confinement) in that it is known to work as it has been empirically demonstrated i.e. there is a strong upper bound on ignition scale, see \cite{Broad1988,Evans2010,Randewich2020}. \cite{Broad1988} suggests this value is approximately 100MJ, the true value may be different. The scale of NIF (2MJ) serves as a lower bound. The uninformative prior for a scale is uniform in log-space (or $\propto \frac{1}{x}$ for a parameter $x$), so for this case study we use a prior uniform in log-space between 2MJ and 100MJ, and zero elsewhere.

\smallskip

We assume in this case that we have one estimate of ignition scale with uncertainty, of $10\pm1$MJ, based on an analysis similar to \cite{Gaffney2019}\footnote{A toy example figure chosen for the purposes of this paper, not a figure based on real analysis of shots.} For both the methods P96 or B18 ideally there would multiple independent estimates of a parameter to begin to constrain the parameters that describe possible unknown systematics, however for this case study we potentially would have only one such estimate. For P96 we can use the previous success rate of predictions for facilities to constrain $\phi$, which can then be used as a prior for this analysis. For this toy example we start with a Jeffery's prior of a $\beta(\frac{1}{2},\frac{1}{2})$ ($p(\theta)=\theta^{\frac{1}{2}}(1-\theta)^{\frac{1}{2}}$) distribution (the non-informative prior for a Bernoulli distribution parameter) as discussed earlier. We then assume we have seen two predictions of ignition proved wrong (NOVA and NIF say) and one good one (correctly predicting SHIVA would not ignite suppose), and update this distribution, and use that as our prior for $\phi$\footnote{Readers may wish to make their own judgements over what facilities to include when forming a prior for success rate in ICF prediction and favour a slightly higher or lower set of values, but to the author the 20-80\% range seems reasonable. }.  It is slightly harder to do something similar for B18 as we don't have multiple simultaneous estimates of ignition scale but we do have historical estimates. Here we use the estimate of $1\pm0.5$ prior to NIF, and combine it with our dummy post-NIF $10\pm1$MJ estimate, and use them to obtain a posterior that we would have found if these estimates had been made simultaneously\footnote{B18 note that their methodology produces unnormalisable posteriors when only two probes are used, unless one has hard priors, so it is fortunate in this case that we have hard upper and lower limits.}. We then shift it to be centred on the post-NIF estimate.

\smallskip

In figure \ref{fig:combined_figure_1} we show the resulting conservative posteriors for ignition scale\footnote{Trying to understand this uncertainty for ICF had an early analysis in section 3b of \cite{Rose2020}.}. In particular it should be noted that even if $\sim$10MJ is most likely, it would still be necessary to build a $\sim$50MJ laser to really achieve certainty that ignition would be achieved.  N.B. \cite{Gaffney2019} allows for some unknown systematics in their analysis, allowing for an unknown source of shot-to-shot variation, so a small amount of unknown unknown systematic is likely allowed for here, but that is more uncertainty at the individual data point, rather than at the probe level. It should also be noted that if a measurement is known to be an underestimate, that is a known unknown, and should be incorporated into the likelihood, P96 and B18 assume that the likelihood are `honest' best efforts at quantification of the uncertainty. P96 and B18 here both allow for unknown unknowns allowing ignition scale to be lower than expected, but past experience might suggest unanticipated physics in ICF normally increases expected ignition scale. In addition, note that Figure \ref{fig:combined_figure_2} shows that the central estimate of the $\phi$ for ICF increases slightly because the single likelihood is consistent with the prior.

\smallskip

ICF experiments are very complex so there are a large number of possible systematics, but possible quantitative systematics include systematics on microphysics like opacity and equation-of-state, or perhaps Rayleigh-Taylor or laser-plasma instabilities being worse than anticipated. Qualitative systematics would correspond to a new phenomenon becoming key that was not previously anticipated e,g. the effects of electrons and ion distributions being non-Maxwellian (if one hadn't anticipated this possibility).

\smallskip

How could independent estimates of ignition scale be calculated, to further constrain $\phi$, ideally so that a higher value of $\phi$ could be justified and more certainty for a given ignition scale claimed? One possible way would be to extrapolate to ignition scale from NIF and Omega shots working independently, using independent simulation codes\footnote{Compare to the recent calls for more multi-team analysis in \cite{Wagenmakers2022}} (HYDRA, \cite{Marinak2001}, and LILAC, \cite{Delettrez1976}, respectively say). It might well be said that the NIF extrapolation will be higher quality and should be trusted more - certainly true. Omega is extrapolating more so the uncertainty on the estimate of ignition scale should be larger. But the degree to which it agrees or not with the NIF extrapolation will constrain to what extent we expect unknown systematic uncertainty; if we estimate $10\pm1$MJ from NIF+Hydra and $10\pm5$MJ from Omega+LILAC, then we have reduced the probability of unknown systematics. If we estimate $10\pm1$MJ from NIF+Hydra and $3\pm5$MJ from Omega+LILAC then clearly there is scope for unknown systematics. Some might say that HYDRA (or LILAC) is simply more accurate than LILAC (or HYDRA) as it incorporates more physics\footnote{Roughly speaking, there is a trade-off between computational expense and inclusion of physics.} - however HYDRA (or LILAC) also will not incorporate all physics, so the size of the discrepancy is key. NIF and Omega use slightly different ignition schemes (indirect and direct respectively) so the described exercise would be non-trivial, but represents one possibility of creating independent estimates. Another complication not discussed here is the possibility of alternative ignition schemes (e.g. Point Design versus Double Shell) - we have implicitly been assuming a 1D parametrisation of the Point Design. Finally we would again note the existence of more radical uncertainty e.g. maybe it is possible to reach ignition at $<$2MJ if a currently unidentified manufacturing issue is affecting current experiments, perhaps ignition does in fact require more than the historically identified upper threshold because the experiments giving that limit were misinterpreted in some way - both unlikely, but possible.

\smallskip

Not from ICF, but we would note that methods that account for unknown unknowns have been applied in making predictions in other areas of nuclear physics. Before the Trinity Test during WWII Manhattan project scientists took bets on what the yield would be. The winning prediction was from Isidor Rabi, who picked 18 kilo-tonnes by simply picking in the middle of the biggest gap with no bet\footnote{According to remarks from Roy Glauber at an Oxford University Physics colloquium, 8th May 2015, \cite{Close2019}. \cite{Rhodes1986} characterises it as 'the only bet left', which may or may not be consistent. Regardless, the winning bet was not chosen for physical reasons.}, essentially taking a uniform prior and picking the value that maximised expected gain given the other bets. In the field of nuclear power plant safety, \cite{Downer2020} claim that `expert knowledge-claims made about catastrophic reactor accidents' normally underestimate risk because reactors normally fail for reasons not previously anticipated (\cite{Ramana2021}) - reactors sometimes claim failure rates of 1 per million years of reactor time, but there have been 15,000 reactor-years to date and more than zero incidents. The incident rate is still likely very small - just much higher than the calculated rate, due to the impact of unexpected phenomena.

\smallskip

Addendum: This sub-section was written before the 2nd August 2021 i.e. prior to the authors knowledge of the campaign leading to the highly successful 8th August 2021 NIF shot\footnote{\fontsize{3.6}{4}{ \url{https://www.llnl.gov/news/national-ignition-facility-experiment-puts-researchers-threshold-fusion-ignition} }}. This shot produced approximately 10 times are much energy as the previous best experiment, and although perhaps not quite reaching full ignition (the full details are still to be published), it likely would suggest the energy required is much lower than some had feared. Would this result surprise someone who had accounted for unknown unknowns? In Figure \ref{fig:combined_figure_1}, although the probe indicated 10MJ, P96, and in particular B18 assign moderate probabilities to values less than 5MJ. Although the 10MJ prediction (and the 100MJ upper limit) was simply chosen to be illustrative, it was comparable with estimates prior to the new result, \cite{LLNL2020}.

\section{Robustness to Malicious Interference}  \label{sec:malicious}

How can one make robust statistical inferences in the presence of the possibility of an unknown third party maliciously interfering with your scientific data? This is not a theoretical concern; there are many cases of hospitals being hit with malicious data breaches\footnote{\url{https://www.theguardian.com/society/2017/may/12/hospitals-across-england-hit-by-large-scale-cyber-attack}}, the Stuxnet computer worm caused centrifuge safety systems to incorrectly give normal data readings\footnote{\url{https://spectrum.ieee.org/the-real-story-of-stuxnet}}, a hacker increased the amount of sodium hydroxide in water supplies in a treatment plant in Florida to dangerous levels\footnote{\fontsize{3.6}{4}{\url{https://www.chemistryworld.com/news/florida-drinking-water-plant-hack-briefly-raised-sodium-hydroxide-levels-100-fold/4013236.article}}}, and the Netherlands Organisation for Scientific Research has been subject to cyberattacks\footnote{\fontsize{3.6}{4}{\url{https://www.sciencemag.org/news/2021/02/dutch-research-funding-agency-paralyzed-ransomware-attack-refuses-pay}}}. \cite{Mirsky2019} showed that generative adversarial networks (GANs) could generate fake medical data that fooled real medical practitioners, and speculated that it could be used to attack individuals, by selectively making it appear that they did/didn't have lung cancer. Scientists working at LIGO detecting gravitational waves (work which was awarded the 2017 Nobel Prize) had genuine concerns (which they ultimately decided were unfounded) at the time that their first detection could be the work of a hoaxer (\cite{Collins2017}).

\smallskip

This risk is also identified in \cite{Dellaportas2020} where Sir Adrian Smith comments \textit{`Well, if you are training machines on data collected by humans, in some sense you have to model the human process that collected it and what the biases and the prejudices and the shortcomings are in that. But, turning this on its head again, could you use mathematics and technology to uncover those biases and correct the learning data sets? In defence and security, depending who you think the good and the bad guys are, what we're all trying to do is subvert the opposition's training data sets by corrupting them. We understand outliers and robustness, so in a scatter plot if I give you a billion concentrated points and one at a distance, a fitted straight line would go through the distant one. Now that's easy to visualise and understand, but what if we've got 10,000 dimensions and a hundred million data points on each, how would you spot that somebody's manipulated a subset in order to disrupt the data set?'} - emphasising the need to model the human process that generated the data.

\smallskip

From the perspective of `unknown unknowns', malicious interference is essentially another source of unmodeled systematic. Treating deliberately incorrect data in a Bayesian manner has also been considered by other authors e.g. \cite{Phillips2021}, and there is an extensive literature of treatment of unknown unknown sensor faults (\cite{Reece2009}). Intuitively P96 seems structured to be more robust to malicious interference\footnote{The text of P96 considers the possibility of `mavericks'} as it can be interpreted as modelling qualitative `wrongness', when considering a scenario of multiple probes of some physics, and an adversary has replaced (unknown to us) one of the probes with data indicating a different value to the true value. P96 also gives a probability of being `wrong' for individual measurements which might help prompt which measurements need further investigation (although of course it doesn't give any indication of whether the discrepancy results from malicious interference or scientific error). 

\smallskip

To test the robustness of inference when using P96 and B18 I generated 100 sets of data with 4 independent probes, and then altered one of the 4 probes, and saw how many times the correct inference was still made. Specifically for each trial I selected a random point uniformly in the range [0,1] to be the `true' value. I then sampled three points from a Gaussian centred on this point with standard deviation 0.05. I then constructed three Gaussians centred on these three points each also with standard deviation 0.05 to be the un-tampered probes. I then constructed one `tampered with' probe by selecting another point uniformly in [0,1], and centring a Gaussian with s.d. 0.05 on top. Counting an estimate as a `success' if the peak of the posterior was within 0.05 of the true value,  CB was correct 40\% of the time, P96 was correct 90\% of the time, and B18 was correct 41\% of the time.  This robustness-check only investigates point estimates, not the realism of the tails, which the rest of this paper focusses on. B18 point estimates were approximately the same as CB but correctly had larger uncertainties. It should also be noted that the P96 method does not just reject `outliers' - the fact that outliers are now known to exist weakens the confidence even in the grouped data points.

\smallskip

\begin{figure*}
\includegraphics[scale=1.0]{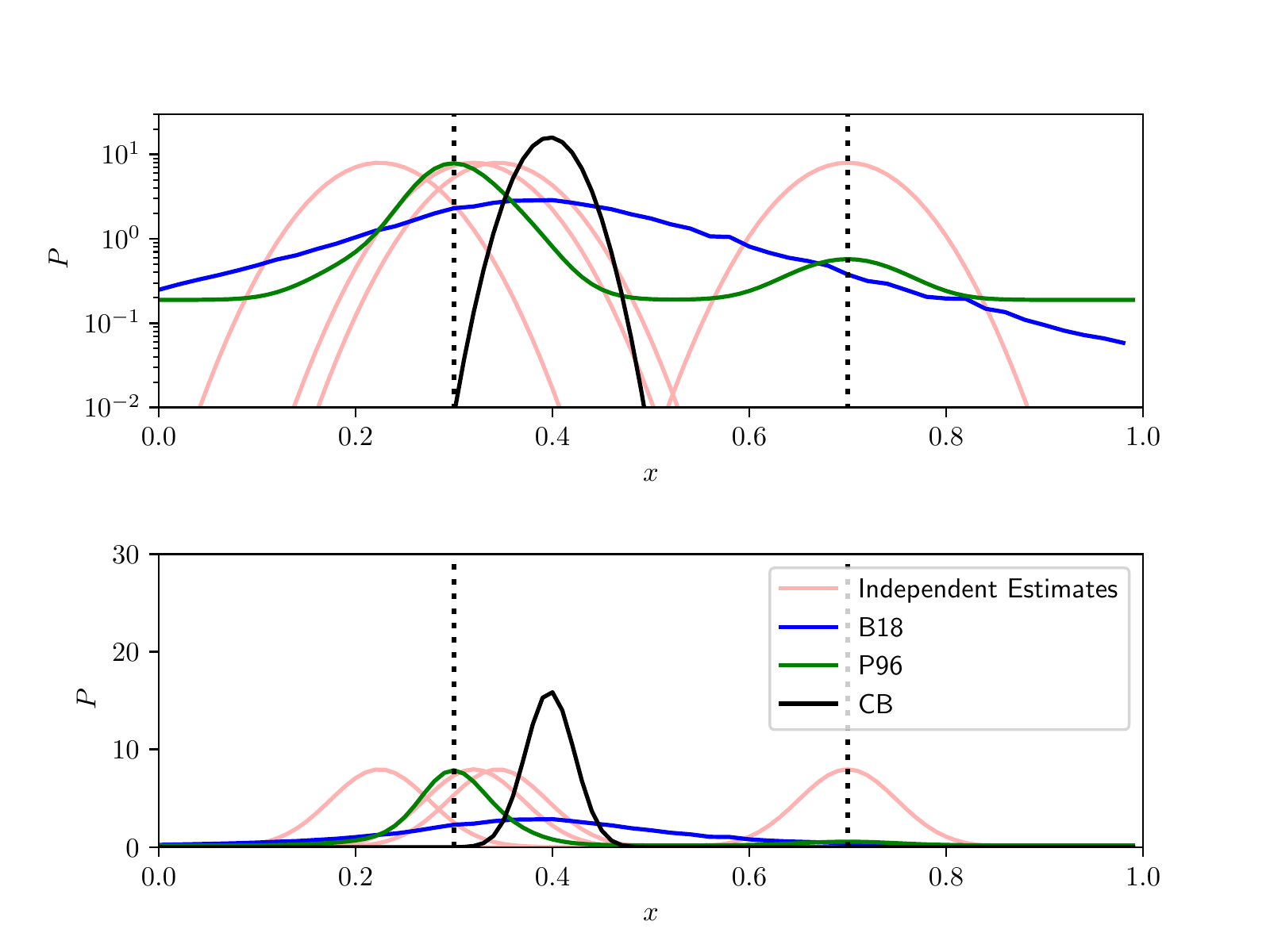}
\caption{Illustrative example of robustness to malicious interference in our simple model of one altered probe. The plots show the Conventional Bayes', P96, B18, and individual probe posteriors (on a log scale for the top sub-plot, and on a linear scale for the lower sub-plot). The dashed line at $x=0.3$ indicates the real value of the parameter, the dashed line at $x=0.7$ indicates where the malicious probe was placed.}
\label{fig:malicious_image}
\end{figure*}

Figure \ref{fig:malicious_image} shows a representative example from the malicious data injection experiment. One can see that P96 is reasonably robust to outliers - this is not too surprising seeing as the data was generated in the exact way that P96 models, but still illustrates the robust nature of the formation of the posterior. To some degree these conservative posteriors can be thought of as error-correcting forms of inference\footnote{Perhaps similar in some regards to error correcting and detecting codes (forms of message where the full message can be recovered even if part of the message is lost e.g. human speech, DNA, some forms of computer memory)}.  Similarly to error correcting codes, the correcting only works up to a point; if a malicious actor was able to corrupt multiple independent sources of data then your inference would eventually be corrupted. We might also note increased generation of synthetic data might make it easier to substitute the `real' data for simulated data. Approaches like this have also been shown to be successful when there is non-malicious interference with data e.g. \cite{Lang2012} use a similar Bayesian model to combine a large number of images of a comet extracted from the internet, most of which really were from the comet, but a small minority were mislabeled images of other objects.

\section{Discussion} \label{sec:discussion_various}

\subsection{Different Forms of Systematic and Correlations}

There are two broad classes of possible systematic present in the above case-studies. Firstly systematics in the data; two groups using the same methodology will come to different conclusions if they have access to different probes each of which might have its own systematics. Secondly systematics in the modelling; two research groups working independently with access to the same data can still come to different conclusions because of different choices and judgements about how to model the data. These are often hard to disentangle and might be ill-defined. But obviously in the second case multiplying liklihoods is probably not justified as they are not independent. In B18 this could be done by placing a prior on the covariance between estimates, in P96 a simple correction might perhaps be to replace products of $N$ liklihoods with the product taken to the $N$th root as done in \cite{Duncan2018a}. 

\smallskip

What constitutes an independent probe is also quite challenging to define. For example in the original problem of cosmological parameter estimation there are multiple independent probes, which disagree to varying degrees. However some of these probes are `early Universe' and some are `late Universe', so it is possible that if the systematics are dependent on whether the probe is late/early, then the systematics will be correlated. Ultimately for true `unknown unknowns' it is intrinsically hard to model the existence of possible correlations like this, but hopefully using a method like P96 or B18 will give better posteriors for making choices even if it is still possible to be `caught out'. In addition P96 and B18 could of course be combined, either heuristically by taking the more conservative posterior in a given scenario, or alternatively by allowing individual probes to be either qualitatively or quantitatively wrong.

\smallskip

One other possible way to make a quasi-independent prediction could be to use machine learning. Machine learning methods typically give biased predictions, but as long as the bias is different to other methods, then they can potentially be used to constrain the size of unknown systematics. \cite{Duncan2018b} use a method similar to P96 to combine theory-based galaxy photometric redshifts with machine learning based empirical galaxy photometric redshifts, giving overall better predictions.

\subsection{Other Possible Approaches to Radical Uncertainty}

P96 and B18 represent just two ways of trying to give realistic posteriors in the presence of unknown systematics. Creating Zipf plots are one alternative approach that contains similar ideas to that presented here. Essentially one would plot a histogram of historical deviances between predicted and true values - in a large set of scenarios measurements like these forms a power law \cite{Clauset2009}\footnote{The importance of power laws in uncertainty quantification is also emphasised in \cite{Taleb2012}}, the index $n$ of which can be measured. This value $n$ could be measured for each situation, which could give a rough indication of how often we might expect unanticipated large systematic discrepancies. Zipf plots are often uses in analysis of distributions of quantities like deaths in wars, where there is no good generative model. Note that many of the P96 and B18 posteriors exhibit power-law like-behaviour.

\smallskip

The key role of unanticipated, low probability, high impact events is noted in the book `The Black Swan' (\cite{Taleb2007}). In it the author identifies 4 quadrants, in a 2x2 matrix, which has `how subject to the unanticipated' on one axis, and `how complex or extreme payoffs can be' on the other axis. The galaxy-halo problem, pair-production and ICF  are likely mainly in the third quadrant, in that they dominated by unanticipated physics, but losses are capped e.g. even if you built a new ICF facility and unanticipated physics prevented it from reaching ignition, the worst that can happen is that it doesn't ignite, and you still learn a multitude of new physics. COVID-19 and pandemics are in the 4th quadrant, where there are many unanticipated effects, and losses are complex and uncapped. For example in the Appendix, deaths per day are capped in the prior at 10,000, but that is not a hard limit, and furthermore we might find that `long covid' effects are more damaging over time etc. More formal methodologies from this area (\cite{Castillo2005,Taleb2020})would be interesting to apply to these problems in future.

\smallskip

In the mathematical literature of Bayesian statistics, a distinction is made between $\mathcal{M}$-closed, $\mathcal{M}$-complete and $\mathcal{M}$-open problems\footnote{There are a range of subtly different definitions in the literature, here we use that of \cite{Clydec2013}} which has some overlap with the issues discussed here, and have a range of treatments, \cite{Hoge2020,Yao2018}. $\mathcal{M}$-closed problems are ones for which there are a finite number of known models, and one of them definitely generated the data. $\mathcal{M}$-complete describes situations where the true model is in some sense `known', but is not in the list of models under consideration because it is too computationally expensive, or has low interpretability, or some other undesirable feature. Finally $\mathcal{M}$-open problems there is no conception of the true model. The galaxy-halo connection and ICF likely fall into $\mathcal{M}$-complete category, in that the models normally used are proxies, but we could in principle do humongous high resolution hydrodynamical cosmological and rad-hydro simulations respectively that would capture all the physics if we had infinite computing power. Pair-production is more of a $\mathcal{M}$-open problem; the true model could be conventional quantum electrodynamics, it could be an axion-based model, or it could be something not yet conceived. Similarly pandemic modelling probably falls into the $\mathcal{M}$-open in that we don't even in principle know how to model everything involved e.g. societal effects. $\mathcal{M}$-closed problems are normally dealt with by summing the predictions of the models, weighted by the implied probabilities from the relevant Bayes factors\footnote{Of course one could also stack the P96 and B18 methods, or do some sort of model comparison between them.}(essentially what is done in \cite{Osthus2019}); $\mathcal{M}$-complete and $\mathcal{M}$-open are more challenging and have a range of approaches in the literature, \cite{Le2017}.

\smallskip

There are a range of other statistical methods and concepts that capture related notions to `unknown unknowns' in different ways to those discussed here. There is the notion of aleatoric versus epistemic uncertainty (\cite{Fox2011}), where in a scientific context aleatoric uncertainty roughly corresponds to stochastic uncertainty, and epistemic uncertainty corresponds to model uncertainty. Our `unknown unknowns' can thus be considered an extreme form of epistemic uncertainty. Not all model uncertainties are `unknown unknowns' however; when measuring a temperature one might know that the thermometer is systematically biased by some unknown amount within some range, a form of model uncertainty - but this is still a `known unknown' as this effect can still be incorporated into the analysis. Then there is the notion of `Fuzzy Logic', which captures some of the idea of allowing something to be partially true, which has some similarities to the notion in P96 of letting a measurement be `good' or `bad' (\cite{Cox1994}). In a machine learning context there is a large literature of work to develop methods for scenarios where the model space doesn't contain a true representation of the data generating process. Methods tackling this problem include using mixtures of Dirichlet processes (\cite{Lyddon2018,Fong2019}). Finally there is also an extensive literature on finding the optimal weightings for combining different neural network predictions, \cite{Penny1999}, and merging hypotheses, \cite{Reece2010}.

\subsection{Practical Use} \label{sec:useful}

The expression `All models are wrong, some are useful.' is also applicable to the construction of our conservative posteriors with P96 and B18. In particular neither P96 and B18 are likely to perfectly represent how unknown systematics in physical systems are generated. More interesting is whether or not these conservative error-correcting posteriors are actually useful. 

\smallskip

It is quite hard to test whether or not these posteriors are more accurate. Firstly because these methods are estimating the probability of rare events, which requires a large number of test cases to compare. Secondly to some degree the purpose of these methods are to identify if there unacceptable risks are being taken (even if of low probability), in which case actions might be taken to mitigate against this possibility (in which case as with the COVID-19 case study in the Appendix, we never get to find out what actually would have happened if no action were taken). One way to determine if posteriors accounting for unknown unknowns were more accurate might be to make blind posterior predictions/calculations for a large number of heterogeneous test cases, and see if the Probability Integral Transform (PIT, e.g. \cite{Dawid1984}) plots were improved.

\smallskip

Another reasonable argument might be that researchers are largely aware of these issues, and that decision makers don't normally take the very low probabilities implied by Gaussian tails literally. When making a decision based on multiple probes normally one might have qualitative expert judgement alongside (e.g. `We know method A and B are highly correlated, that is why they are giving the same value', `Method C is known to be very unreliable' etc.). In response to this it might be argued a) that some of this expert judgement is still known unknowns, and should be incorporated to the modelling (e.g. if it is known that Method C is poor it should be given a very large uncertainty) and b) these conservative error-correcting posteriors are attempting to give indicative order of magnitude estimates of how likely extreme deviations are (as opposed to exact statistics), which might be hard to do qualitatively. Finally we would reiterate the comment in the Introduction that what qualifies as a good posterior is typically highly dependent on the problem, which typically demand different statistical properties - some problems might have high costs to extreme results, others might not.

\smallskip

It might also fairly be commented that these methods are uninformative in that they do not tell you the physical source of discrepancy. P96 and B18 don't tell you what the cause of the fat tails is - they are just saying that models aren't agreeing strongly enough to justify claiming that you have ruled out unknown unknowns. The methods also still require a moderate amount of judgement e.g. choosing what probes to count as independent probes, construction of prior on $\phi$, P96 versus B18. In response to this issue I would again comment that this is not `precision statistics', more an attempt to get tail probabilities correct to an order of magnitude, up to a few orders of magnitudes away from the bulk of the data.

\smallskip

The issues discussed here are also potentially relevant to the use of statistics in the presentation of scientific evidence in other scenarios e.g. perhaps in court (where a trial is of course a form of decision making), \cite{Evett2016}. For example DNA evidence historically was presented with claims of very high levels of reliability (along the lines that there was `only a 1 in ten billion chance the sample not from the defendant' etc.). This may well be the probability of an incorrect match of two DNA samples, but what is the probability of two samples being mixed up, or someone working in the lab being involved in the case and tampering with the data and so on ad infinitum? Quite low, but likely more probable than 1 in ten billion, so in that regard calculating the probability of scientific consistency is sub-dominant to the role of unknown unknowns\footnote{Note that having multiple `probes' is also considered by some to be beneficial for robust inference in the context of court cases, \cite{Sangero2007}.} that could impact the data.

\section{Conclusion} \label{sec:conclusion}

Conventional application of Bayes' Theorem gives good estimates of confidence, but still underestimates the true uncertainty as it doesn't account for the possibility of unknown systematics, both in data, and in modelling. In particular this usually leads to underestimates of the thickness of posterior tails. This is often appreciated, but not often quantified. In this paper we discussed two methods in the literature to attempt to get more realistic uncertainties (even though the possible existence of rare unknown unknowns is intrinsically hard to model), and applied them to a series of case studies. The Galaxy-Halo Connection case study illustrated a scenario where we had separate probes of some underlying physics, but the probes likely would still have some correlated systematics, as the probes both came from the same instrument. The Matter from Light case illustrated a scenario where the probes really were highly independent, but risked becoming so different that the underlying physics wasn't comparable. Finally the ICF case study illustrated an example where only one probe was available. We have shown the exact shape and size of tail risks depends on if qualitative or quantitative unknowns are dominant. We also speculate that P96 or similar approaches might be useful to mitigate against malicious interference of data.  If one is in a scenario where unknown systematics are likely to be mainly of quantitive physical origin B18 may be better, if there is a reasonable fear of deliberate interference P96 might be more informative.\\
\\
\textbf{Key Conclusions:}
\begin{itemize}
	\item Hierarchical Bayesian models that incorporate the probability of `being wrong' can be used to give more realistic posteriors
	\item Qualitative and quantitive unknown unknowns can be modelled in different ways
	\item In the presence of unknown systematics, different probes and models should be kept as far apart as possible to maintain independence
	\item A form of error-correcting physical inference should be used in situations where there are risks scientific data could be tampered with
	\item These attempts to model unknown systematics are still just models and thus will still fail in some circumstances
\end{itemize}

\section{Acknowledgements}

PH acknowledges support from a GCHQ Research Fellowship for National Resilience for this work. This paper received valuable input from Professor Steven Rose (Imperial College London), and Professor Matt Jarvis, Dr Aprajita Verma, Professor David Alonso, and Professor Stephen Roberts (University of Oxford).

\vspace{1cm}

\appendix


Although not an astronomy or physics example, in the course of preparing this manuscript the events of the coronavirus pandemic in the UK did present an opportunity to test the calculation of conservative posteriors in a real life emergency. The author would note that they have no epidemiological or medical expertise, so this Appendix should be regarded as a toy example, as opposed to a serious comment on the COVID-19 pandemic. Similarly the author claims no particular insight into how the decisions were actually made (for example how much weight was given to different models in the formation of the scientific advice, and to what degree the probability of tail risks was qualitatively known). In particular however this toy model was  implemented `blind' before the fact (on 10th November 2020), so the analysis does constitute an attempt at deploying these methods in a semi-realistic scenario\footnote{What is presented here is a slightly updated analysis, original 2020 analysis available on request.}. Bayesian ensembling methods have been used successfully by other authors in the literature for COVID-19 forecasting, \cite{Adiga2021}.

\smallskip

During the COVID-19 pandemic governments have used epidemiological models (e.g. \cite{Moore2021}) to try and understand how many deaths might be expected under certain scenarios, under what timeframes, to what degree of confidence. These models typically take large quantities of data (number of positive tests etc.) as input and again use some model fitting process (often a variant on Bayes' theorem) to make predictions by running the model forwards in time. Different models typically don't perfectly agree, include different assumptions and rely on different sets of data etc. A particular decision making point we focus on is the decision to go into the second English lockdown, announced 31st October 2020, lasting from the 5th November 2020 to 2nd December 2020. This was based to some degree on the predictions from four models, from Imperial, Warwick, Public Health England (PHE)/Cambridge and The London School of Hygiene \& Tropical Medicine (LSHTM)\footnote{\url{https://assets.publishing.service.gov.uk/government/uploads/system/uploads/attachment_data/file/932251/Slides_presented_by_Chief_Scientific_Advisor_to_accompany_coronavirus_press_conference_31_October_2020.pdf}}. The Imperial, Warwick and LSHTM models predicted deaths peaking at about 2,000 a day, with the PHE/Cambridge model predicting higher numbers, around 4,000\footnote{It might be commented that total deaths are the more important figure, but for simplicity here we focus on peak number of deaths per day.}. All predicted this peak occurring late December/early January. However on the 5th November 2020 the UK Statistics Authority criticised the inclusion of that version of the PHE/Cambridge model in decision making on the grounds that it was known to be based on slightly out of date data\footnote{\url{https://www.bbc.co.uk/news/health-54831334}} (known to me when I did my initial analysis). However, over the winter of 2020/2021 the situation changed dramatically with the emergence of the `Kent' or `UK' variant (B.1.1.7, or `Alpha') of the virus, a mutated form that spread much faster (\cite{Davies2021}), identified to be the cause of faster spread of the virus on approximately the 21st December 2020 (i.e. unknown to me at time of my blind analysis). The emergence of the new variant was essentially an unknown unknown that presumably would have rendered many of the assumptions in the models incorrect and changed predictions by more than quoted uncertainties\footnote{The possibility of variants had always been known as a possibility, but to the authors understanding it was not known at the time that variants would be the main way in which the pandemic would evolve.}.  In reality deaths peaked at about $\sim1,400$ deaths per day around the 20th January 2021 - although this cannot be directly compared to the predictions as the models were attempting to describe what would happen if no further interventions were taken.

\smallskip

\begin{figure}
\includegraphics[scale=0.6]{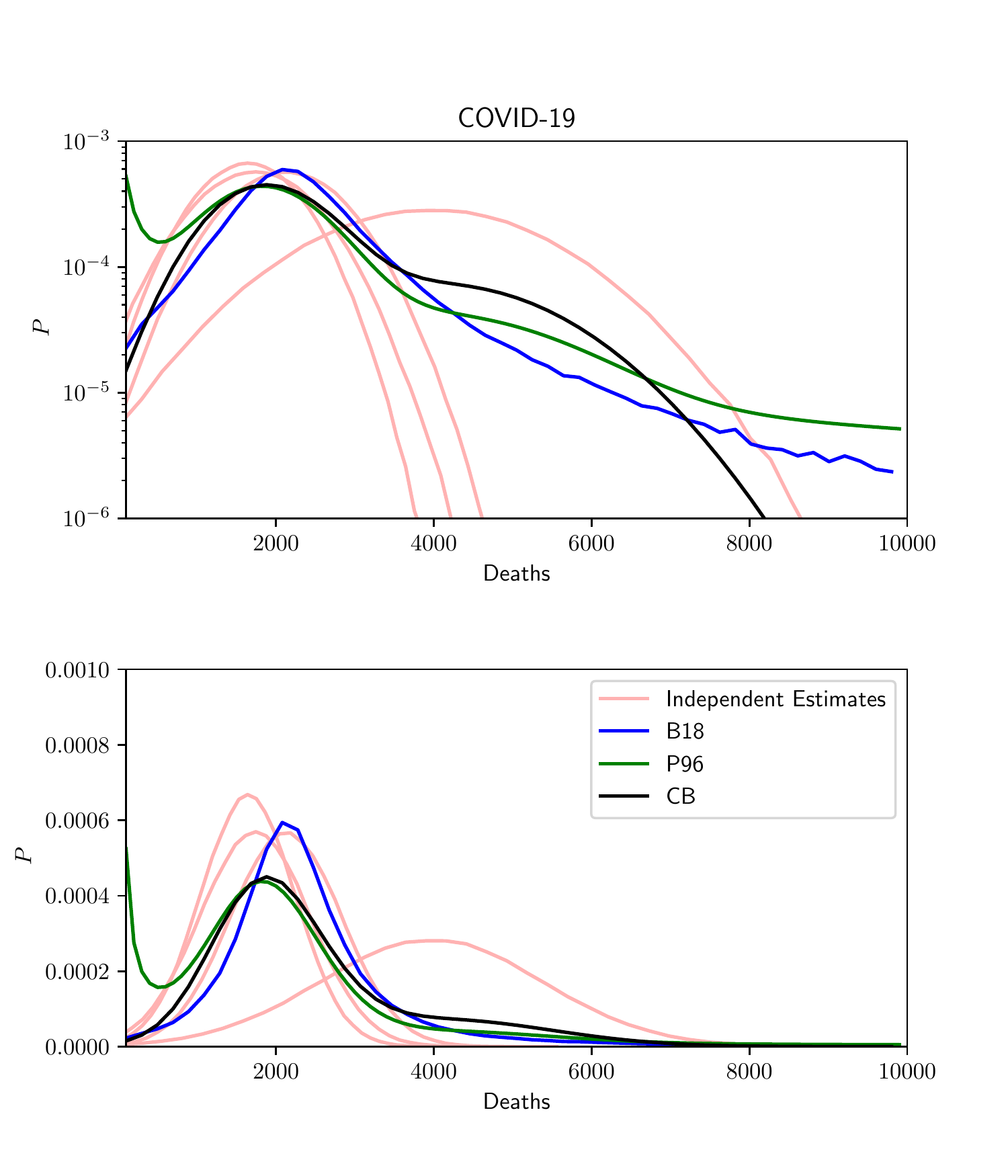}
\caption{The posteriors for the coronavirus case study. Probabilities are shown in both log (top) and log space (bottom). Black shows the conventional posterior, green shows the P96 posterior, blue shows the B18 and the light red are the distributions from the individual separate models.}
\label{fig:covid_combined_figure_1}
\end{figure}

\begin{figure}
\includegraphics[scale=0.6]{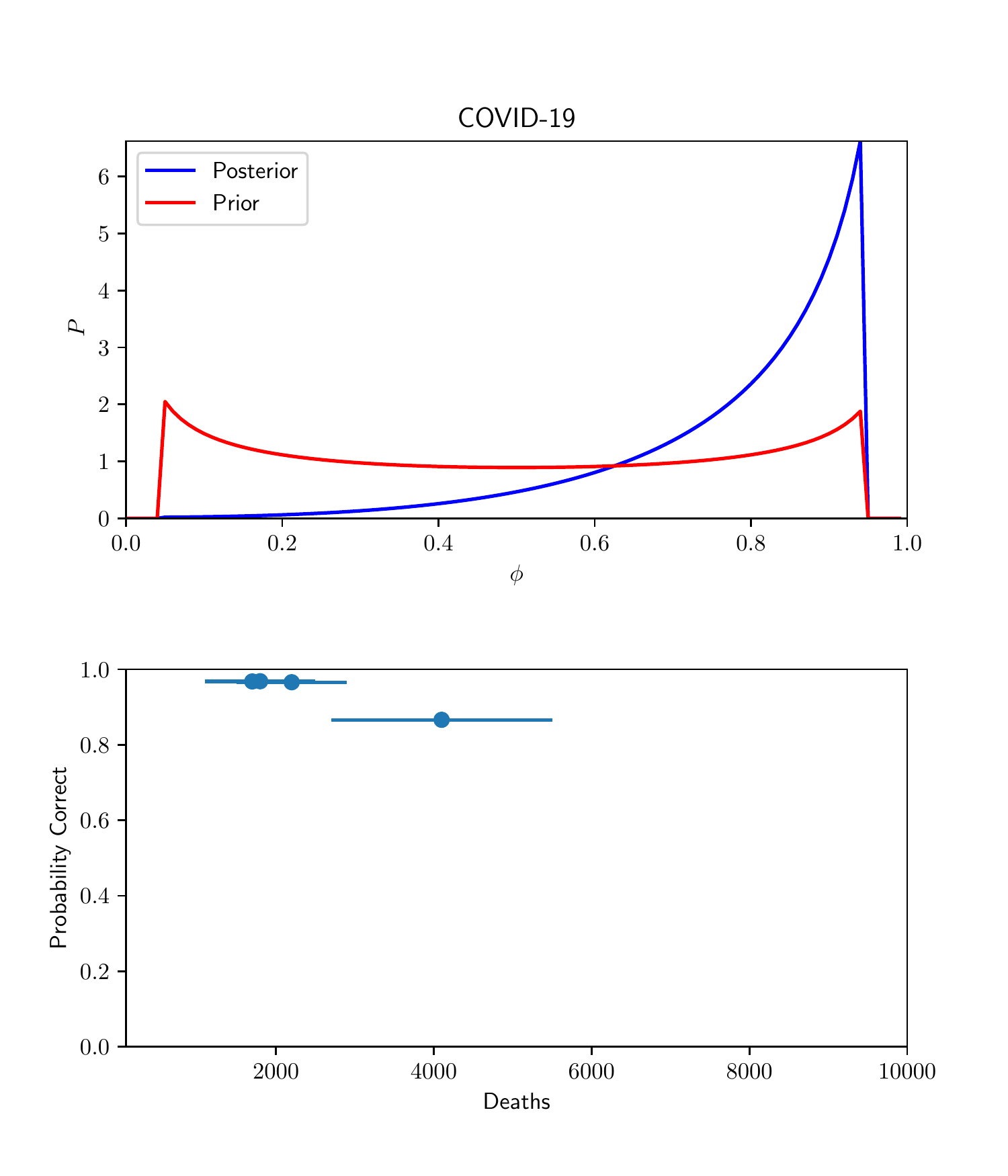}
\caption{The parameters describing the probabilities of measurements being correct in the P96 model for the coronavirus case study. The top sub-figure shows the global probabilities (the red curves corresponding to the prior, and the blue curves to the posterior), the lower sub-plot shows the probabilities for individual models.}
\label{fig:covid_combined_figure_2}
\end{figure}

This scenario is slightly different to the earlier case studies in that the different estimates are not clear distinct probes but rather a more complex mix of different models (which take different but overlapping input data). We therefore slightly modify our analysis. For Conventional Bayes we instead take the mixture model of the four models (equal weighting, simple model averaging), and for P96 we restrict it to only allow one of the models to be true at a time (e.g. a mixture model of the four models and the prior, with weights as per P96). Rigorously modifying B18 would require a more complex understanding of the covariances between the models (potentially quantified with a bootstrapping method), but for the illustrative purposes here we replace the pdf from each probe with its $N$th root (c.f. \cite{Dahlen2013,Duncan2018a}).  In the analysis I used predicted peak deaths per day of $1700\pm600$, $1800\pm700$, $2200\pm700$ and $4100\pm1400]$\footnote{These were approximate values read off the SAGE graph, as opposed to the actual likelihoods from the modelling, so are not exact}, results shown in Figure \ref{fig:covid_combined_figure_1} and \ref{fig:covid_combined_figure_2}. Both P96 and B18 give fat tails to higher numbers of deaths, although a peak of around 2,000 remains the most probable value in the model. In particular they give non-trivial probabilities to numbers beyond the contribution in CB from the PHE/Cambridge model, giving some small probability for mechanisms beyond those imagined in any model. The probes are fairly consistent, so the estimate of $\phi$ increases. It is hard to definitively prove that P96 and B18 better capture the uncertainties as the distributions make a prediction for a one-off event, and we can never measure what would have happened with no interventions. However hopefully the reader will agree in the light of unanticipated events like B.1.1.7 that even worse unexpected events (not necessarily a worse mutation - winter weather combined with COVID might have had much worse than anticipated combined effects, national oxygen supplies for hospitals might have run out, the unrelated chicken bird flu taking place in the UK at the same time\footnote{\url{https://www.theguardian.com/environment/2020/dec/14/new-measures-begin-to-help-curb-british-bird-flu-cases-in-poultry}} might have crossed to humans and caused greater mortality when an individual had both COVID-19 and bird flu...) that might have pushed deaths up to even higher levels were possible. New phenomena like variants might count as qualitative unknown unknowns; something like the virus incubation period being shorter than believed might count as a quantitative unknown unknown. It might also be noted that the emergency nature of a pandemic potentially makes the risk of human error in assembling data that is incorporated into inference greater e.g. statistics about tests lost\footnote{\url{https://www.bbc.co.uk/news/technology-54423988}}. Doctors in the BMA have commented that several aspects of the evolution of the pandemic were unanticipated\footnote{\url{https://www.bmj.com/content/371/bmj.m3979}} and others have emphasised the role of `unknown unknowns', \cite{Luo2021}. The fat-tailed nature of pandemic risk has been emphasised by some authors e.g. \cite{Cirillo2020} point out that damage from pandemics likely follows a power-law distribution. Finally it is important to note that the PHE/Cambridge estimate that differed from the other estimates did not predict the cause of the unknown effect (the new variant), it differed for other reasons. Instead P96 and B18 are using the fact that the quasi-independent methods (to which here we are agnostic about which to prefer) did not completely agree to constrain how likely possible severe unanticipated deviations are.

\bibliographystyle{aasjournal}

\clearpage

\bibliography{paper}

\begin{thebibliography}{}
\expandafter\ifx\csname natexlab\endcsname\relax\def\natexlab#1{#1}\fi
\providecommand{\url}[1]{\href{#1}{#1}}
\providecommand{\dodoi}[1]{doi:~\href{http://doi.org/#1}{\nolinkurl{#1}}}
\providecommand{\doeprint}[1]{\href{http://ascl.net/#1}{\nolinkurl{http://ascl.net/#1}}}
\providecommand{\doarXiv}[1]{\href{https://arxiv.org/abs/#1}{\nolinkurl{https://arxiv.org/abs/#1}}}

\bibitem[{Abbas {et~al.}(2013)Abbas, Abelev, Adam, Adamov{\'{a}}, Adare,
  Aggarwal, {Aglieri Rinella}, Agnello, Agocs, Agostinelli, Ahammed, Ahmad,
  {Ahmad Masoodi}, Ahmed, Ahn, Ahn, Aimo, Ajaz, Akindinov, Aleksandrov,
  Alessandro, Alexandre, Alici, Alkin, {Almar{\'{a}}z Avi{\~{n}}a}, Alme, Alt,
  Altini, Altinpinar, Altsybeev, Andrei, Andronic, Anguelov, Anielski, Anson,
  Anti{\v{c}}i{\'{c}}, Antinori, Antonioli, Aphecetche, Appelsh{\"{a}}user,
  Arbor, Arcelli, Arend, Armesto, Arnaldi, Aronsson, Arsene, Arslandok, Asryan,
  Augustinus, Averbeck, Awes, {\"{A}}yst{\"{o}}, Azmi, Bach, Badal{\`{a}},
  Baek, Bailhache, Bala, Baldisseri, {Baltasar Dos Santos Pedrosa}, B{\'{a}}n,
  Baral, Barbera, Barile, Barnaf{\"{o}}ldi, Barnby, Barret, Bartke, Basile,
  Bastid, Basu, Bathen, Batigne, Batyunya, Batzing, Baumann, Bearden, Beck,
  Behera, Belikov, Bellini, Bellwied, Belmont-Moreno, Bencedi, Beole, Berceanu,
  Bercuci, Berdnikov, Berenyi, Bergognon, Bertens, Berzano, Betev, Bhasin,
  Bhati, Bhom, Bianchi, Bianchi, Bianchin, Biel{\v{c}}{\'{i}}k,
  Biel{\v{c}}{\'{i}}kov{\'{a}}, Bilandzic, Bjelogrlic, Blanco, Blanco, Blau,
  Blume, Boccioli, B{\"{o}}ttger, Bogdanov, B{\o}ggild, Bogolyubsky,
  Boldizs{\'{a}}r, Bombara, Book, Borel, Borissov, Boss{\'{u}}, Botje, Botta,
  Braidot, Braun-Munzinger, Bregant, Breitner, Broker, Browning, Broz, Brun,
  Bruna, Bruno, Budnikov, Buesching, Bufalino, Buncic, Busch, Buthelezi,
  Caffarri, Cai, Caines, {Calvo Villar}, Camerini, {Canoa Roman}, {Cara Romeo},
  Carena, Carena, {Carlin Filho}, Carminati, {Casanova D{\'{i}}az}, {Castillo
  Castellanos}, {Castillo Hernandez}, Casula, Catanescu, Cavicchioli, {Ceballos
  Sanchez}, Cepila, Cerello, Chang, Chapeland, Charvet, Chattopadhyay,
  Chattopadhyay, Cherney, Cheshkov, Cheynis, {Chibante Barroso}, Chinellato,
  Chochula, Chojnacki, Choudhury, Christakoglou, Christensen, Christiansen,
  Chujo, Chung, Cicalo, Cifarelli, Cindolo, Cleymans, Colamaria, Colella,
  Collu, {Conesa Balbastre}, {Conesa del Valle}, Connors, Contin, Contreras,
  Cormier, {Corrales Morales}, Cortese, {Cort{\'{e}}s Maldonado}, Cosentino,
  Costa, Cotallo, Crescio, Crochet, {Cruz Alaniz}, {Cruz Albino}, Cuautle,
  Cunqueiro, Dainese, Dang, Danu, Das, Das, Das, Das, Dash, Dash, De,
  de~Barros, {De Caro}, de~Cataldo, de~Cuveland, {De Falco}, {De Gruttola},
  Delagrange, Deloff, {De Marco}, D{\'{e}}nes, {De Pasquale}, Deppman, {D
  Erasmo}, de~Rooij, {Diaz Corchero}, {Di Bari}, Dietel, {Di Giglio}, {Di
  Liberto}, {Di Mauro}, {Di Nezza}, Divi{\`{a}}, Djuvsland, Dobrin,
  Dobrowolski, D{\"{o}}nigus, Dordic, Dubey, Dubla, Ducroux, Dupieux, {Dutta
  Majumdar}, Elia, Emschermann, Engel, Erazmus, Erdal, Eschweiler, Espagnon,
  Estienne, Esumi, Evans, Evdokimov, Eyyubova, Fabris, Faivre, Falchieri,
  Fantoni, Fasel, Fehlker, Feldkamp, Felea, Feliciello, Fenton-Olsen, Feofilov,
  {Fern{\'{a}}ndez T{\'{e}}llez}, Ferretti, Festanti, Figiel, Figueredo,
  Filchagin, Finogeev, Fionda, Fiore, Floratos, Floris, Foertsch, Foka, Fokin,
  Fragiacomo, Francescon, Frankenfeld, Fuchs, Furget, {Fusco Girard},
  Gaardh{\o}je, Gagliardi, Gago, Gallio, Gangadharan, Ganoti, Garabatos,
  Garcia-Solis, Gargiulo, Garishvili, Gerhard, Germain, Geuna, Gheata, Gheata,
  Ghidini, Ghosh, Gianotti, Giubellino, Gladysz-Dziadus, Gl{\"{a}}ssel, Gomez,
  Ferreiro, Gonz{\'{a}}lez-Trueba, Gonz{\'{a}}lez-Zamora, Gorbunov, Goswami,
  Gotovac, Graczykowski, Grajcarek, Grelli, Grigoras, Grigoras, Grigoriev,
  Grigoryan, Grigoryan, Grinyov, Grion, Gros, Grosse-Oetringhaus, Grossiord,
  Grosso, Guber, Guernane, Guerzoni, Guilbaud, Gulbrandsen, Gulkanyan, Gunji,
  Gupta, Gupta, Haake, Haaland, Hadjidakis, Haiduc, Hamagaki, Hamar, Han,
  Hanratty, Hansen, Harmanov{\'{a}}-T{\'{o}}thov{\'{a}}, Harris, Hartig,
  Harton, Hatzifotiadou, Hayashi, Hayrapetyan, Heckel, Heide, Helstrup,
  Herghelegiu, {Herrera Corral}, Herrmann, Hess, Hetland, Hicks, Hippolyte,
  Hori, Hristov, Hřivn{\'{a}}{\v{c}}ov{\'{a}}, Huang, Humanic, Hwang, Ichou,
  Ilkaev, Ilkiv, Inaba, Incani, Innocenti, Innocenti, Ippolitov, Irfan, Ivan,
  Ivanov, Ivanov, Ivanov, Ivanytskyi, Jacho{\l}kowski, Jacobs, Jahnke, Jang,
  Janik, Jayarathna, Jena, Jha, {Jimenez Bustamante}, Jones, Jung, Jusko,
  Kaidalov, Kalcher, Kaliň{\'{a}}k, Kalliokoski, Kalweit, Kang, Kaplin, Kar,
  {Karasu Uysal}, Karavichev, Karavicheva, Karpechev, Kazantsev, Kebschull,
  Keidel, Ketzer, Khan, Khan, Khan, Khan, Khanzadeev, Kharlov, Kileng, Kim,
  Kim, Kim, Kim, Kim, Kim, Kim, Kim, Kim, Kirsch, Kisel, Kiselev, Kisiel, Klay,
  Klein, Klein-B{\"{o}}sing, Kliemant, Kluge, Knichel, Knospe, K{\"{o}}hler,
  Kollegger, Kolojvari, Kompaniets, Kondratiev, Kondratyeva, Konevskikh,
  Kovalenko, Kowalski, Kox, {Koyithatta Meethaleveedu}, Kral, Kr{\'{a}}lik,
  Kramer, Krav{\v{c}}{\'{a}}kov{\'{a}}, Krelina, Kretz, Krivda, Krizek, Krus,
  Kryshen, Krzewicki, Kucera, Kucheriaev, Kugathasan, Kuhn, Kuijer, Kulakov,
  Kumar, Kurashvili, Kurepin, Kurepin, Kuryakin, Kushpil, Kushpil, Kvaerno,
  Kweon, Kwon, {Ladr{\'{o}}n de Guevara}, {Lagana Fernandes}, Lakomov, Langoy,
  {La Pointe}, Lara, Lardeux, {La Rocca}, Lea, Lechman, Lee, Lee, Legrand,
  Lehnert, Lemmon, Lenhardt, Lenti, Le{\'{o}}n, Leoncino, {Le{\'{o}}n
  Monz{\'{o}}n}, L{\'{e}}vai, Li, Lien, Lietava, Lindal, Lindenstruth,
  Lippmann, Lisa, Ljunggren, Lodato, Loenne, Loggins, Loginov, Lohner,
  Loizides, Loo, Lopez, {L{\'{o}}pez Torres}, L{\o}vh{\o}iden, Lu, Luettig,
  Lunardon, Luo, Luparello, Luzzi, Ma, Ma, Madagodahettige-Don, Maevskaya,
  Mager, Mahapatra, Maire, Malaev, {Maldonado Cervantes}, Malinina, Mal'Kevich,
  Malzacher, Mamonov, Manceau, Mangotra, Manko, Manso, Manzari, Mao,
  Marchisone, Mare{\v{s}}, Margagliotti, Margotti, Mar{\'{i}}n, Markert,
  Marquard, Martashvili, Martin, Martinengo, Mart{\'{i}}nez, {Mart{\'{i}}nez
  Garc{\'{i}}a}, Martynov, Mas, Masciocchi, Masera, Masoni, Massacrier,
  Mastroserio, Matyja, Mayer, Mazer, Mazumder, Mazzoni, Meddi, Menchaca-Rocha,
  {Mercado P{\'{e}}rez}, Meres, Miake, Mikhaylov, Milano, Milosevic, Mischke,
  Mishra, Mi{\'{s}}kowiec, Mitu, Mizuno, Mlynarz, Mohanty, Molnar,
  {Monta{\~{n}}o Zetina}, Monteno, Montes, Moon, Morando, {Moreira De Godoy},
  Moretto, Morreale, Morsch, Muccifora, Mudnic, Muhuri, Mukherjee,
  M{\"{u}}ller, Munhoz, Murray, Musa, Musinsky, Nandi, Nania, Nappi, Nattrass,
  Nayak, Nazarenko, Nedosekin, Nicassio, Niculescu, Nielsen, Niida, Nikolaev,
  Nikolic, Nikulin, Nikulin, Nilsen, Nilsson, Noferini, Nomokonov, Nooren,
  Nyanin, Nyatha, Nygaard, Nystrand, Ochirov, Oeschler, Oh, Oh, Oleniacz,
  {Oliveira Da Silva}, Onderwaater, Oppedisano, {Ortiz Velasquez}, Oskarsson,
  Ostrowski, Otwinowski, Oyama, Ozawa, Pachmayer, Pachr, Padilla, Pagano,
  Pai{\'{c}}, Painke, Pajares, Pal, Palaha, Palmeri, Papikyan, Pappalardo,
  Park, Passfeld, Patalakha, Paticchio, Paul, Pavlinov, Pawlak, Peitzmann,
  {Pereira Da Costa}, {Pereira De Oliveira Filho}, Peresunko, {P{\'{e}}rez
  Lara}, Perrino, Peryt, Pesci, Pestov, Petr{\'{a}}{\v{c}}ek, Petran, Petris,
  Petrov, Petrovici, Petta, Piano, Pikna, Pillot, Pinazza, Pinsky, Pitz,
  Piyarathna, Planinic, P{\l}osko{\'{n}}, Pluta, Pocheptsov, Pochybova,
  Podesta-Lerma, Poghosyan, Pol{\'{a}}k, Polichtchouk, Poljak, Pop,
  Porteboeuf-Houssais, Posp{\'{i}}{\v{s}}il, Potukuchi, Prasad, Preghenella,
  Prino, Pruneau, Pshenichnov, Puddu, Punin, Putschke, Qvigstad, Rachevski,
  Rademakers, R{\"{a}}ih{\"{a}}, Rak, Rakotozafindrabe, Ramello, Raniwala,
  Raniwala, R{\"{a}}s{\"{a}}nen, Rascanu, Rathee, Rauch, Rauf, Razazi, Read,
  Real, Redlich, Reed, Rehman, Reichelt, Reicher, Renfordt, Reolon, Reshetin,
  Rettig, Revol, Reygers, Riccati, Ricci, Richert, Richter, Riedler, Riegler,
  Riggi, Rivetti, {Rodr{\'{i}}guez Cahuantzi}, {Rodriguez Manso}, R{\o}ed,
  Rogochaya, Rohr, R{\"{o}}hrich, Romita, Ronchetti, Rosnet, Rossegger, Rossi,
  Roy, Roy, {Rubio Montero}, Rui, Russo, Ryabinkin, Rybicki, Sadovsky,
  {\v{S}}afař{\'{i}}k, Sahoo, Sahu, Saini, Sakaguchi, Sakai, Sakata, Salgado,
  Salzwedel, Sambyal, Samsonov, {Sanchez Castro}, {\v{S}}{\'{a}}ndor, Sandoval,
  Sano, Santagati, Santoro, Sarkamo, Sarkar, Scapparone, Scarlassara,
  Scharenberg, Schiaua, Schicker, Schmidt, Schmidt, Schuchmann, Schukraft,
  Schuster, Schutz, Schwarz, Schweda, Scioli, Scomparin, Scott, Scott, Segato,
  Selyuzhenkov, Senyukov, Seo, Serci, Serradilla, Sevcenco, Shabetai,
  Shabratova, Shahoyan, Sharma, Sharma, Rohni, Shigaki, Shtejer, Sibiriak,
  Sicking, Siddhanta, Siemiarczuk, Silvermyr, Silvestre, Simatovic, Simonetti,
  Singaraju, Singh, Singha, Singhal, Sinha, Sinha, Sitar, Sitta, Skaali,
  Skjerdal, Smakal, Smirnov, Snellings, S{\o}gaard, Soltz, Song, Song, Soos,
  Soramel, Sputowska, Spyropoulou-Stassinaki, Srivastava, Stachel, Stan,
  Stefanek, Steinpreis, Stenlund, Steyn, Stiller, Stocco, Stolpovskiy, Strmen,
  Suaide, {Subieta V{\'{a}}squez}, Sugitate, Suire, Suleymanov, Sultanov,
  {\v{S}}umbera, Susa, Symons, {Szanto de Toledo}, Szarka, Szczepankiewicz,
  Szyma{\'{n}}ski, Takahashi, Tangaro, {Tapia Takaki}, {Tarantola Peloni},
  {Tarazona Martinez}, Tauro, {Tejeda Mu{\~{n}}oz}, Telesca, {Ter Minasyan},
  Terrevoli, Th{\"{a}}der, Thomas, Tieulent, Timmins, Tlusty, Toia, Torii,
  Toscano, Trubnikov, Truesdale, Trzaska, Tsuji, Tumkin, Turrisi, Tveter,
  Ulery, Ullaland, Ulrich, Uras, Urciuoli, Usai, Vajzer, Vala, {Valencia
  Palomo}, Vallero, {Vande Vyvre}, {Van Hoorne}, van Leeuwen, Vannucci, Vargas,
  Varma, Vasileiou, Vasiliev, Vechernin, Veldhoen, Venaruzzo, Vercellin,
  Vergara, Vernet, Verweij, Vickovic, Viesti, Viinikainen, Vilakazi,
  {Villalobos Baillie}, Vinogradov, Vinogradov, Vinogradov, Virgili, Viyogi,
  Vodopyanov, V{\"{o}}lkl, Voloshin, Voloshin, Volpe, von Haller, Vorobyev,
  Vranic, Vrl{\'{a}}kov{\'{a}}, Vulpescu, Vyushin, Wagner, Wagner, Wan, Wang,
  Wang, Wang, Watanabe, Weber, Wessels, Westerhoff, Wiechula, Wikne, Wilde,
  Wilk, Williams, Windelband, Yaldo, Yamaguchi, Yang, Yang, Yang, Yasnopolskiy,
  Yi, Yin, Yoo, Yoon, Yuan, Yushmanov, Zaccolo, Zach, Zampolli, Zaporozhets,
  Zarochentsev, Z{\'{a}}vada, Zaviyalov, Zbroszczyk, Zelnicek, Zgura, Zhalov,
  Zhang, Zhang, Zhang, Zhou, Zhou, Zhou, Zhu, Zhu, Zhu, Zhu, Zichichi,
  Zimmermann, Zinovjev, Zoccarato, Zynovyev, \& Zyzak}]{Abbas2013}
Abbas, E., Abelev, B., Adam, J., {et~al.} 2013, The European Physical Journal
  C, 73, 2617, \dodoi{10.1140/epjc/s10052-013-2617-1}

\bibitem[{Adam {et~al.}(2012)Adam, Agafonova, Aleksandrov, Altinok, Sanchez,
  Anokhina, Aoki, Ariga, Ariga, Autiero, Badertscher, Dhahbi, Bertolin, Bozza,
  Brugi{\`{e}}re, Brugnera, Brunet, Brunetti, Buontempo, Carlus, Cavanna,
  Cazes, Chaussard, Chernyavsky, Chiarella, Chukanov, Colosimo, Crespi,
  D'Ambrosio, {De Lellis}, {De Serio}, D{\'{e}}clais, {Del Amo Sanchez}, {Di
  Capua}, {Di Crescenzo}, {Di Ferdinando}, {Di Marco}, Dmitrievsky, Dracos,
  Duchesneau, Dusini, Dzhatdoev, Ebert, Efthymiopoulos, Egorov, Ereditato,
  Esposito, Favier, Ferber, Fini, Fukuda, Garfagnini, Giacomelli, Giorgini,
  Giovannozzi, Girerd, Goldberg, G{\"{o}}llnitz, Golubkov, Goncharova,
  Gornushkin, Grella, Grianti, Gschwendtner, Guerin, Guler, Gustavino, Hagner,
  Hamada, Hara, Enikeev, Hierholzer, Hollnagel, Ieva, Ishida, Ishiguro,
  Jakovcic, Jollet, Jones, Juget, Kamiscioglu, Kawada, Kim, Kimura, Kiritsis,
  Kitagawa, Klicek, Knuesel, Kodama, Komatsu, Kose, Kreslo, Lazzaro, Lenkeit,
  Ljubicic, Longhin, Malgin, Mandrioli, Marteau, Matsuo, Matveev, Mauri,
  Mazzoni, Medinaceli, Meisel, Meregaglia, Migliozzi, Mikado, Missiaen,
  Monacelli, Morishima, Moser, Muciaccia, Naganawa, Naka, Nakamura, Nakano,
  Nakatsuka, Naumov, Nikitina, Nitti, Ogawa, Okateva, Olchevsky, Palamara,
  Paoloni, Park, Park, Pastore, Patrizii, Pennacchio, Pessard, Pistillo,
  Polukhina, Pozzato, Pretzl, Pupilli, Rescigno, Riguzzi, Roganova, Rokujo,
  Rosa, Rostovtseva, Rubbia, Russo, Ryasny, Ryazhskaya, Sato, Sato, Sahnoun,
  Schembri, Schuler, Lavina, Serrano, Shakiryanova, Sheshukov, Shibuya,
  Shoziyoev, Simone, Sioli, Sirignano, Sirri, Song, Spinetti, Stanco, Starkov,
  Stellacci, Stipcevic, Strauss, Takahashi, Tenti, Terranova, Tezuka, Tioukov,
  Tolun, Tran, Tufanli, Vilain, Vladimirov, Votano, Vuilleumier, Wilquet,
  Wonsak, Wurtz, Yakushev, Yoon, Yoshida, Zaitsev, Zemskova, \&
  Zghiche}]{Adam2012}
Adam, T., Agafonova, N., Aleksandrov, A., {et~al.} 2012, Journal of High Energy
  Physics, 2012, 93, \dodoi{10.1007/JHEP10(2012)093}

\bibitem[{Adiga {et~al.}(2021)Adiga, Wang, Hurt, Peddireddy, Porebski,
  Venkatramanan, Lewis, \& Marathe}]{Adiga2021}
Adiga, A., Wang, L., Hurt, B., {et~al.} 2021, medRxiv,
  \dodoi{10.1101/2021.03.12.21253495}

\bibitem[{Andrae(2010)}]{Andrae2010}
Andrae, R. 2010.
\newblock \doarXiv{1009.2755}

\bibitem[{Barlow(2002)}]{Barlow2002}
Barlow, R. 2002.
\newblock \doarXiv{0207026}

\bibitem[{Bernal {et~al.}(2018)Bernal, Peacock, Bernal, \&
  Peacock}]{Bernal2018}
Bernal, J.~L., Peacock, J.~A., Bernal, J.~L., \& Peacock, J.~A. 2018, JCAP,
  2018, 002, \dodoi{10.1088/1475-7516/2018/07/002}

\bibitem[{Biteau \& Williams(2015)}]{Biteau2015}
Biteau, J., \& Williams, D.~A. 2015, The Astrophysical Journal, 812, 60,
  \dodoi{10.1088/0004-637X/812/1/60}

\bibitem[{Breit \& Wheeler(1934)}]{Breit1934}
Breit, G., \& Wheeler, J.~A. 1934, Physical Review, 46, 1087,
  \dodoi{10.1103/PhysRev.46.1087}

\bibitem[{Broad(1988)}]{Broad1988}
Broad, W.~G. 1988, {Secret Advance in Nuclear Fusion Spurs a Dispute Among
  Scientists, New York Times}.
\newblock
  \url{https://www.nytimes.com/1988/03/21/us/secret-advance-in-nuclear-fusion-spurs-a-dispute-among-scientists.html}

\bibitem[{Burke {et~al.}(1997)Burke, Field, Horton-Smith, Spencer, Walz,
  Berridge, Bugg, Shmakov, Weidemann, Bula, {Mc Donald}, Prebys, Bamber, Boege,
  Koffas, Kotseroglou, Melissinos, Meyerhofer, Reis, \& Ragg}]{Burke1997}
Burke, D.~L., Field, R.~C., Horton-Smith, G., {et~al.} 1997, Physical Review
  Letters, 79, 1626, \dodoi{10.1103/PhysRevLett.79.1626}

\bibitem[{Camporeale(2019)}]{Camporeale2019a}
Camporeale, E. 2019, Space Weather, 17, 1166, \dodoi{10.1029/2018SW002061}

\bibitem[{Castillo {et~al.}(2005)Castillo, Hadi, Sarabia, \&
  Balakrishnan}]{Castillo2005}
Castillo, E., Hadi, A., Sarabia, J.~M., \& Balakrishnan, N. 2005, International
  Statistical Institute, 368.
\newblock \url{http://www.lavoisier.fr/notice/frJWO3LRLAXRWLKO.html}

\bibitem[{Cirillo \& Taleb(2020)}]{Cirillo2020}
Cirillo, P., \& Taleb, N.~N. 2020, Nature Physics, 16, 606,
  \dodoi{10.1038/s41567-020-0921-x}

\bibitem[{Clauset {et~al.}(2009)Clauset, Shalizi, \& Newman}]{Clauset2009}
Clauset, A., Shalizi, C.~R., \& Newman, M.~E. 2009, SIAM Review, 51, 661.
\newblock \doarXiv{0706.1062}

\bibitem[{Close(2019)}]{Close2019}
Close, F. 2019, {Trinity} (Penguin Books)

\bibitem[{Clydec \& Iversen(2013)}]{Clydec2013}
Clydec, M., \& Iversen, E.~S. 2013, Bayesian Theory and Applications, 484,
  \dodoi{10.1093/ACPROF:OSO/9780199695607.003.0024}

\bibitem[{Collins(2017)}]{Collins2017}
Collins, H. 2017, {Gravity's Kiss} (MIT Press)

\bibitem[{Coupon {et~al.}(2015)Coupon, Arnouts, van Waerbeke, Moutard, Ilbert,
  van Uitert, Erben, Garilli, Guzzo, Heymans, Hildebrandt, Hoekstra, Kilbinger,
  Kitching, Mellier, Miller, Scodeggio, Bonnett, Branchini, Davidzon, {De
  Lucia}, Fritz, Fu, Hudelot, Hudson, Kuijken, Leauthaud, {Le Fevre},
  McCracken, Moscardini, Rowe, Schrabback, Semboloni, \& Velander}]{Coupon2015}
Coupon, J., Arnouts, S., van Waerbeke, L., {et~al.} 2015, Monthly Notices of
  the Royal Astronomical Society, 449, 1352, \dodoi{10.1093/mnras/stv276}

\bibitem[{Cowen(2015)}]{Cowen2015}
Cowen, R. 2015, Nature, \dodoi{10.1038/nature.2015.16830}

\bibitem[{Cox(1994)}]{Cox1994}
Cox, E. 1994, {The fuzzy systems handbook : a practitioner's guide to building
  and maintaining fuzzy systems} (AP Professional), 615

\bibitem[{Dahlen {et~al.}(2013)Dahlen, Mobasher, Faber, Ferguson, Barro,
  Finkelstein, Finlator, Fontana, Gruetzbauch, Johnson, Pforr, Salvato,
  Wiklind, Wuyts, Acquaviva, Dickinson, Guo, Huang, Huang, Newman, Bell,
  Conselice, Galametz, Gawiser, Giavalisco, Grogin, Hathi, Kocevski, Koekemoer,
  Koo, Lee, McGrath, Papovich, Peth, Ryan, Somerville, Weiner, \&
  Wilson}]{Dahlen2013}
Dahlen, T., Mobasher, B., Faber, S.~M., {et~al.} 2013, The Astrophysical
  Journal, 775, 93, \dodoi{10.1088/0004-637X/775/2/93}

\bibitem[{Davies {et~al.}(2021)Davies, Abbott, Barnard, Jarvis, Kucharski,
  Munday, Pearson, Russell, Tully, Washburne, Wenseleers, Gimma, Waites, Wong,
  van Zandvoort, Silverman, Group1‡, Consortium‡, Diaz-Ordaz, Keogh, Eggo,
  Funk, Jit, Atkins, \& Edmunds}]{Davies2021}
Davies, N.~G., Abbott, S., Barnard, R.~C., {et~al.} 2021, Science, 372,
  eabg3055, \dodoi{10.1126/SCIENCE.ABG3055}

\bibitem[{Dawid(1984)}]{Dawid1984}
Dawid, A.~P. 1984, Journal of the Royal Statistical Society. Series A
  (General), 147, 278, \dodoi{10.2307/2981683}

\bibitem[{Delettrez \& Goldman(1976)}]{Delettrez1976}
Delettrez, J., \& Goldman, E.~B. 1976, {Laboratory for Laser Energetics Report
  No. 36, University of Rochester. See also National Technical In- formation
  Service. Document No. DOE/SF/19460-118.}, Tech. rep.

\bibitem[{Dellaportas \& Stephens(2020)}]{Dellaportas2020}
Dellaportas, P., \& Stephens, D.~A. 2020, International Statistical Review, 88,
  265, \dodoi{10.1111/insr.12395}

\bibitem[{{Di Valentino} {et~al.}(2021){Di Valentino}, Mena, Pan, Visinelli,
  Yang, Melchiorri, Mota, Riess, \& Silk}]{DiValentino2021}
{Di Valentino}, E., Mena, O., Pan, S., {et~al.} 2021, Classical and Quantum
  Gravity, 38, 153001, \dodoi{10.1088/1361-6382/ac086d}

\bibitem[{Downer \& Ramana(2020)}]{Downer2020}
Downer, J., \& Ramana, M.~V. 2020, Regulation {\&} Governance, rego.12300,
  \dodoi{10.1111/rego.12300}

\bibitem[{Draper(1995)}]{Draper1995}
Draper, D. 1995, Journal of the Royal Statistical Society: Series B
  (Methodological), 57, 45, \dodoi{10.1111/j.2517-6161.1995.tb02015.x}

\bibitem[{Draper {et~al.}(1987)Draper, Hodges, Leamer, Morris, \&
  Rubin}]{Draper1987}
Draper, D., Hodges, J.~S., Leamer, E.~E., Morris, C.~N., \& Rubin, D.~B. 1987.
\newblock \url{https://www.rand.org/pubs/notes/N2683.html}

\bibitem[{Duncan {et~al.}(2018{\natexlab{a}})Duncan, Jarvis, Brown, \&
  R{\"{o}}ttgering}]{Duncan2018b}
Duncan, K.~J., Jarvis, M.~J., Brown, M. J.~I., \& R{\"{o}}ttgering, H. J.~A.
  2018{\natexlab{a}}, Monthly Notices of the Royal Astronomical Society, 477,
  5177, \dodoi{10.1093/mnras/sty940}

\bibitem[{Duncan {et~al.}(2018{\natexlab{b}})Duncan, Brown, Williams, Best,
  Buat, Burgarella, Jarvis, Ma{\l}ek, Oliver, R{\"{o}}ttgering, Smith, Malek,
  Oliver, R{\"{o}}ttgering, \& Smith}]{Duncan2018a}
Duncan, K.~J., Brown, M. J.~I., Williams, W.~L., {et~al.} 2018{\natexlab{b}},
  Monthly Notices of the Royal Astronomical Society, 473, 2655,
  \dodoi{10.1093/mnras/stx2536}

\bibitem[{Evans {et~al.}(2019)Evans, O'Hare, \& McCabe}]{Evans2019}
Evans, N.~W., O'Hare, C.~A., \& McCabe, C. 2019, Physical Review D, 99, 023012,
  \dodoi{10.1103/PhysRevD.99.023012}

\bibitem[{Evans(2010)}]{Evans2010}
Evans, R. 2010, Physics World, 23, 23, \dodoi{10.1088/2058-7058/23/03/34}

\bibitem[{Evett {et~al.}(2016)Evett, Pope, \& Puch-Solis}]{Evett2016}
Evett, I., Pope, S., \& Puch-Solis, R. 2016, Science {\&} Justice, 56, 278,
  \dodoi{10.1016/j.scijus.2016.05.001}

\bibitem[{Feeney {et~al.}(2018)Feeney, Mortlock, \& Dalmasso}]{Feeney2018}
Feeney, S.~M., Mortlock, D.~J., \& Dalmasso, N. 2018, Monthly Notices of the
  Royal Astronomical Society, 476, 3861, \dodoi{10.1093/mnras/sty418}

\bibitem[{Fong {et~al.}(2019)Fong, Lyddon, \& Holmes}]{Fong2019}
Fong, E., Lyddon, S., \& Holmes, C. 2019, {Scalable Nonparametric Sampling from
  Multimodal Posteriors with the Posterior Bootstrap},  PMLR.
\newblock \url{http://proceedings.mlr.press/v97/fong19a.html}

\bibitem[{Futter(2018)}]{Futter2018}
Futter, A. 2018, {Hacking the Bomb} (Georgetown University Press)

\bibitem[{Gaffney {et~al.}(2019)Gaffney, Brandon, Humbird, Kruse, Nora,
  Peterson, \& Spears}]{Gaffney2019}
Gaffney, J.~A., Brandon, S.~T., Humbird, K.~D., {et~al.} 2019, Physics of
  Plasmas, 26, 082704, \dodoi{10.1063/1.5108667}

\bibitem[{Gelman {et~al.}(2014)Gelman, Carlin, Stern, \& Rubin}]{Gelman2014}
Gelman, A., Carlin, J. B.~B., Stern, H. S.~S., \& Rubin, D. B.~B. 2014, Book,
  675

\bibitem[{Grabo(2002)}]{Grabo2002}
Grabo, C. 2002, {Anticipating Surprise: Analysis for Strategic Warning}.
\newblock
  \url{https://rowman.com/ISBN/9780761829522/Anticipating-Surprise-Analysis-for-Strategic-Warning}

\bibitem[{Haidich(2010)}]{Haidich2010}
Haidich, A.~B. 2010, Hippokratia, 14, 29.
\newblock \url{/pmc/articles/PMC3049418/
  /pmc/articles/PMC3049418/?report=abstract
  https://www.ncbi.nlm.nih.gov/pmc/articles/PMC3049418/}

\bibitem[{Hall {et~al.}(2021)Hall, Davies, van Saders, Nielsen, Lund, Chaplin,
  Garc{\'{i}}a, Amard, Breimann, Khan, See, \& Tayar}]{Hall2021}
Hall, O.~J., Davies, G.~R., van Saders, J., {et~al.} 2021, Nature Astronomy, 5,
  707, \dodoi{10.1038/s41550-021-01335-x}

\bibitem[{Hammer {et~al.}(2005)Hammer, Bildsten, Abarbanel, Cornwall, Eardley,
  Happer, Flatt{\'{e}}, Hemley, Jeanloz, Katz, Max, Meiron, Perkins, Prentiss,
  Schwitters, Bodner, Gardner, Hoffer, \& Program}]{Hammer2005}
Hammer, D.~A., Bildsten, L., Abarbanel, H., {et~al.} 2005

\bibitem[{Hatfield {et~al.}(2016)Hatfield, Lindsay, Jarvis, H{\"{a}}u{\ss}ler,
  Vaccari, \& Verma}]{Hatfield2016}
Hatfield, P.~W., Lindsay, S.~N., Jarvis, M.~J., {et~al.} 2016, Monthly Notices
  of the Royal Astronomical Society, 459, 2618, \dodoi{10.1093/mnras/stw769}

\bibitem[{Hatfield {et~al.}(2019)Hatfield, Rose, \& Scott}]{Hatfield2019}
Hatfield, P.~W., Rose, S.~J., \& Scott, R. H.~H. 2019, Physics of Plasmas, 26,
  062706, \dodoi{10.1063/1.5091985}

\bibitem[{Hatfield {et~al.}(2021)Hatfield, Gaffney, Anderson, Ali, Antonelli,
  {Başeğmez du Pree}, Citrin, Fajardo, Knapp, Kettle, Kustowski, MacDonald,
  Mariscal, Martin, Nagayama, Palmer, Peterson, Rose, Ruby, Shneider, Streeter,
  Trickey, \& Williams}]{Hatfield2021}
Hatfield, P.~W., Gaffney, J.~A., Anderson, G.~J., {et~al.} 2021, Nature, 593,
  351, \dodoi{10.1038/s41586-021-03382-w}

\bibitem[{Hinton {et~al.}(2019)Hinton, Davis, Kim, Brout, D'Andrea, Kessler,
  Lasker, Lidman, Macaulay, M{\"{o}}ller, Sako, Scolnic, Smith, Wolf,
  Childress, Morganson, Allam, Annis, Avila, Bertin, Brooks, Burke, Rosell,
  Kind, Carretero, Cunha, da~Costa, Davis, Vicente, DePoy, Doel, Eifler,
  Flaugher, Fosalba, Frieman, Garc{\'{i}}a-Bellido, Gaztanaga, Gerdes, Gruendl,
  Gschwend, Gutierrez, Hartley, Hollowood, Honscheid, Krause, Kuehn,
  Kuropatkin, Lahav, Lima, Maia, March, Marshall, Menanteau, Miquel, Ogando,
  Plazas, Sanchez, Scarpine, Schindler, Schubnell, Serrano, Sevilla-Noarbe,
  Soares-Santos, Sobreira, Suchyta, Tarle, Thomas, Vikram, \&
  Zhang}]{Hinton2019}
Hinton, S.~R., Davis, T.~M., Kim, A.~G., {et~al.} 2019, The Astrophysical
  Journal, 876, 15, \dodoi{10.3847/1538-4357/ab13a3}

\bibitem[{Hobson {et~al.}(2002)Hobson, Bridle, \& Lahav}]{Hobson2002}
Hobson, M.~P., Bridle, S.~L., \& Lahav, O. 2002, Monthly Notices of the Royal
  Astronomical Society, 335, 377, \dodoi{10.1046/j.1365-8711.2002.05614.x}

\bibitem[{H{\"{o}}ge {et~al.}(2020)H{\"{o}}ge, Guthke, \& Nowak}]{Hoge2020}
H{\"{o}}ge, M., Guthke, A., \& Nowak, W. 2020, Water 2020, Vol. 12, Page 309,
  12, 309, \dodoi{10.3390/W12020309}

\bibitem[{Horns {et~al.}(2012)Horns, MacCione, Meyer, Mirizzi, Montanino, \&
  Roncadelli}]{Horns2012}
Horns, D., MacCione, L., Meyer, M., {et~al.} 2012, Physical Review D -
  Particles, Fields, Gravitation and Cosmology, 86, 075024,
  \dodoi{10.1103/PhysRevD.86.075024}

\bibitem[{Huang(2010)}]{Huang2010}
Huang, Q. 2010, IIE Transactions, 43, 1, \dodoi{10.1080/07408171003795335}

\bibitem[{Hurricane {et~al.}(2016)Hurricane, Callahan, Casey, Dewald, Dittrich,
  D{\"{o}}ppner, Haan, Hinkel, {Berzak Hopkins}, Jones, Kritcher, {Le Pape},
  Ma, MacPhee, Milovich, Moody, Pak, Park, Patel, Ralph, Robey, Ross,
  Salmonson, Spears, Springer, Tommasini, Albert, Benedetti, Bionta, Bond,
  Bradley, Caggiano, Celliers, Cerjan, Church, Dylla-Spears, Edgell, Edwards,
  Fittinghoff, {Barrios Garcia}, Hamza, Hatarik, Herrmann, Hohenberger, Hoover,
  Kline, Kyrala, Kozioziemski, Grim, Field, Frenje, Izumi, {Gatu Johnson},
  Khan, Knauer, Kohut, Landen, Merrill, Michel, Moore, Nagel, Nikroo, Parham,
  Rygg, Sayre, Schneider, Shaughnessy, Strozzi, Town, Turnbull, Volegov, Wan,
  Widmann, Wilde, \& Yeamans}]{Hurricane2016a}
Hurricane, O.~A., Callahan, D.~A., Casey, D.~T., {et~al.} 2016, Nature Physics,
  12, 800, \dodoi{10.1038/nphys3720}

\bibitem[{Ishikawa {et~al.}(2020)Ishikawa, Kashikawa, Tanaka, Coupon,
  Leauthaud, Toshikawa, Ichikawa, Oogi, Uchiyama, Niino, Nishizawa, Ishikawa,
  Kashikawa, Tanaka, Coupon, Leauthaud, Toshikawa, Ichikawa, Oogi, Uchiyama,
  Niino, \& Nishizawa}]{Ishikawa2020}
Ishikawa, S., Kashikawa, N., Tanaka, M., {et~al.} 2020, ApJ, 904, 128,
  \dodoi{10.3847/1538-4357/ABBD95}

\bibitem[{Ivezi{\'{c}} {et~al.}(2019)Ivezi{\'{c}}, Kahn, Tyson, Abel, Acosta,
  Allsman, Alonso, AlSayyad, Anderson, Andrew, {P. Angel}, Angeli, Ansari,
  Antilogus, Araujo, Armstrong, Arndt, Astier, Aubourg, Auza, Axelrod, Bard,
  Barr, Barrau, Bartlett, Bauer, Bauman, Baumont, Bechtol, Bechtol, Becker,
  Becla, Beldica, Bellavia, Bianco, Biswas, Blanc, Blazek, Blandford, Bloom,
  Bogart, Bond, Booth, Borgland, Borne, Bosch, Boutigny, Brackett, Bradshaw,
  Brandt, Brown, Bullock, Burchat, Burke, Cagnoli, Calabrese, Callahan, Callen,
  Carlin, Carlson, Chandrasekharan, Charles-Emerson, Chesley, Cheu, Chiang,
  Chiang, Chirino, Chow, Ciardi, Claver, Cohen-Tanugi, Cockrum, Coles,
  Connolly, Cook, Cooray, Covey, Cribbs, Cui, Cutri, Daly, Daniel, Daruich,
  Daubard, Daues, Dawson, Delgado, Dellapenna, de~Peyster, de~Val-Borro, Digel,
  Doherty, Dubois, Dubois-Felsmann, Durech, Economou, Eifler, Eracleous,
  Emmons, Neto, Ferguson, Figueroa, Fisher-Levine, Focke, Foss, Frank, Freemon,
  Gangler, Gawiser, Geary, Gee, Geha, Gessner, Gibson, Gilmore, Glanzman,
  Glick, Goldina, Goldstein, Goodenow, Graham, Gressler, Gris, Guy, Guyonnet,
  Haller, Harris, Hascall, Haupt, Hernandez, Herrmann, Hileman, Hoblitt,
  Hodgson, Hogan, Howard, Huang, Huffer, Ingraham, Innes, Jacoby, Jain, Jammes,
  Jee, Jenness, Jernigan, Jevremovi{\'{c}}, Johns, Johnson, Johnson, Jones,
  Juramy-Gilles, Juri{\'{c}}, Kalirai, Kallivayalil, Kalmbach, Kantor, Karst,
  Kasliwal, Kelly, Kessler, Kinnison, Kirkby, Knox, Kotov, Krabbendam,
  Krughoff, Kub{\'{a}}nek, Kuczewski, Kulkarni, Ku, Kurita, Lage, Lambert,
  Lange, Langton, Guillou, Levine, Liang, Lim, Lintott, Long, Lopez, Lotz,
  Lupton, Lust, MacArthur, Mahabal, Mandelbaum, Markiewicz, Marsh, Marshall,
  Marshall, May, McKercher, McQueen, Meyers, Migliore, Miller, Mills, Miraval,
  Moeyens, Moolekamp, Monet, Moniez, Monkewitz, Montgomery, Morrison, Mueller,
  Muller, Arancibia, Neill, Newbry, Nief, Nomerotski, Nordby, O'Connor, Oliver,
  Olivier, Olsen, O'Mullane, Ortiz, Osier, Owen, Pain, Palecek, Parejko,
  Parsons, Pease, Peterson, Peterson, Petravick, Petrick, Petry, Pierfederici,
  Pietrowicz, Pike, Pinto, Plante, Plate, Plutchak, Price, Prouza, Radeka,
  Rajagopal, Rasmussen, Regnault, Reil, Reiss, Reuter, Ridgway, Riot, Ritz,
  Robinson, Roby, Roodman, Rosing, Roucelle, Rumore, Russo, Saha, Sassolas,
  Schalk, Schellart, Schindler, Schmidt, Schneider, Schneider, Schoening,
  Schumacher, Schwamb, Sebag, Selvy, Sembroski, Seppala, Serio, Serrano, Shaw,
  Shipsey, Sick, Silvestri, Slater, Smith, Smith, Sobhani, Soldahl,
  Storrie-Lombardi, Stover, Strauss, Street, Stubbs, Sullivan, Sweeney,
  Swinbank, Szalay, Takacs, Tether, Thaler, Thayer, Thomas, Thornton, Thukral,
  Tice, Trilling, Turri, Berg, Berk, Vetter, Virieux, Vucina, Wahl, Walkowicz,
  Walsh, Walter, Wang, Wang, Warner, Wiecha, Willman, Winters, Wittman, Wolff,
  Wood-Vasey, Wu, Xin, Yoachim, \& Zhan}]{Ivezic2019}
Ivezi{\'{c}}, {\v{Z}}., Kahn, S.~M., Tyson, J.~A., {et~al.} 2019, The
  Astrophysical Journal, 873, 111, \dodoi{10.3847/1538-4357/ab042c}

\bibitem[{Jackman(2009)}]{Jackman2009}
Jackman, S. 2009, {Bayesian Analysis for the Social Sciences} (Wiley)

\bibitem[{Jarvis {et~al.}(2013)Jarvis, Bonfield, Bruce, Geach, McAlpine,
  McLure, Gonzalez-Solares, Irwin, Lewis, Yoldas, Andreon, Cross, Emerson,
  Dalton, Dunlop, Hodgkin, Le, Karouzos, Meisenheimer, Oliver, Rawlings,
  Simpson, Smail, Smith, Sullivan, Sutherland, White, \& Zwart}]{Jarvis2013}
Jarvis, M.~J., Bonfield, D.~G., Bruce, V.~A., {et~al.} 2013, Monthly Notices of
  the Royal Astronomical Society, 428, 1281, \dodoi{10.1093/mnras/sts118}

\bibitem[{Jeffreys(1946)}]{JEFFREYS1946}
Jeffreys, H. 1946, Proceedings of the Royal Society of London. Series A.
  Mathematical and Physical Sciences, 186, 453, \dodoi{10.1098/rspa.1946.0056}

\bibitem[{Kasim {et~al.}(2019)Kasim, Galligan, Topp-Mugglestone, Gregori, \&
  Vinko}]{Kasim2019}
Kasim, M.~F., Galligan, T.~P., Topp-Mugglestone, J., Gregori, G., \& Vinko,
  S.~M. 2019, Physics of Plasmas, 26, 112706, \dodoi{10.1063/1.5125979}

\bibitem[{Katzfuss {et~al.}(2017)Katzfuss, Hammerling, \& Smith}]{Katzfuss2017}
Katzfuss, M., Hammerling, D., \& Smith, R.~L. 2017, Geophysical Research
  Letters, 44, 5720, \dodoi{10.1002/2017GL073688}

\bibitem[{Kennedy \& O'Hagan(2001)}]{Kennedy2001}
Kennedy, M.~C., \& O'Hagan, A. 2001, Journal of the Royal Statistical Society:
  Series B (Statistical Methodology), 63, 425, \dodoi{10.1111/1467-9868.00294}

\bibitem[{Kettle {et~al.}(2021)Kettle, Hollatz, Gerstmayr, Samarin, Alejo,
  Astbury, Baird, Bohlen, Campbell, Colgan, Dannheim, Gregory, Harsh, Hatfield,
  Hinojosa, Katzir, Morton, Murphy, Nurnberg, Osterhoff, P{\'{e}}rez-Callejo,
  P{\~{o}}der, Rajeev, Roedel, Roeder, Salgado, Sarri, Seidel, Spannagel,
  Spindloe, Steinke, Streeter, Thomas, Underwood, Watt, Zepf, Rose, \&
  Mangles}]{Kettle2021}
Kettle, B., Hollatz, D., Gerstmayr, E., {et~al.} 2021, New Journal of Physics,
  23, 115006, \dodoi{10.1088/1367-2630/ac3048}

\bibitem[{K{\"{o}}hlinger {et~al.}(2019)K{\"{o}}hlinger, Joachimi, Asgari,
  Viola, Joudaki, \& Tr{\"{o}}ster}]{Kohlinger2019}
K{\"{o}}hlinger, F., Joachimi, B., Asgari, M., {et~al.} 2019, Monthly Notices
  of the Royal Astronomical Society, 484, 3126, \dodoi{10.1093/mnras/stz132}

\bibitem[{Lang \& Hogg(2012)}]{Lang2012}
Lang, D., \& Hogg, D.~W. 2012, The Astronomical Journal, 144, 46,
  \dodoi{10.1088/0004-6256/144/2/46}

\bibitem[{Le \& Clarke(2017)}]{Le2017}
Le, T., \& Clarke, B. 2017, Bayesian Analysis, 12, 807,
  \dodoi{10.1214/16-BA1023}

\bibitem[{{Le Cozannet} {et~al.}(2020){Le Cozannet}, Kervyn, Russo, {Ifejika
  Speranza}, Ferrier, Foumelis, Lopez, \& Modaressi}]{LeCozannet2020}
{Le Cozannet}, G., Kervyn, M., Russo, S., {et~al.} 2020, {Space-Based Earth
  Observations for Disaster Risk Management},  Springer Science and Business
  Media B.V., \dodoi{10.1007/s10712-020-09586-5}

\bibitem[{Li {et~al.}(2016)Li, Chen, Jiang, Apley, Lu, \& Chen}]{Li2016}
Li, W., Chen, S., Jiang, Z., {et~al.} 2016, Journal of Verification, Validation
  and Uncertainty Quantification, 1, \dodoi{10.1115/1.4031983}

\bibitem[{Lindl(1995)}]{Lindl1995}
Lindl, J. 1995, Physics of Plasmas, 2, 3933, \dodoi{10.1063/1.871025}

\bibitem[{Lindl {et~al.}(2004)Lindl, Amendt, Berger, Glendinning, Glenzer,
  Haan, Kauffman, Landen, \& Suter}]{Lindl2004}
Lindl, J.~D., Amendt, P., Berger, R.~L., {et~al.} 2004, Physics of Plasmas, 11,
  339, \dodoi{10.1063/1.1578638}

\bibitem[{LLNL(2020)}]{LLNL2020}
LLNL. 2020, {Laser Indirect Drive input to NNSA 2020 Report}, Tech. rep.

\bibitem[{Luo(2021)}]{Luo2021}
Luo, J. 2021, Technological forecasting and social change, 166,
  \dodoi{10.1016/J.TECHFORE.2021.120602}

\bibitem[{Lyddon {et~al.}(2018)Lyddon, Walker, \& Holmes}]{Lyddon2018}
Lyddon, S., Walker, S., \& Holmes, C.~C. 2018, Advances in Neural Information
  Processing Systems, 31

\bibitem[{MacMahon {et~al.}(2004)MacMahon, Pearce, \& Harris}]{MacMahon2004}
MacMahon, D., Pearce, A., \& Harris, P. 2004, Applied Radiation and Isotopes,
  60, 275, \dodoi{10.1016/j.apradiso.2003.11.028}

\bibitem[{Marinak {et~al.}(2001)Marinak, Kerbel, Gentile, Jones, Munro,
  Pollaine, Dittrich, \& Haan}]{Marinak2001}
Marinak, M.~M., Kerbel, G.~D., Gentile, N.~A., {et~al.} 2001, Physics of
  Plasmas, 8, 2275, \dodoi{10.1063/1.1356740}

\bibitem[{Mirsky {et~al.}(2019)Mirsky, Mahler, Shelef, \& Elovici}]{Mirsky2019}
Mirsky, Y., Mahler, T., Shelef, I., \& Elovici, Y. 2019, in Proceedings of the
  28th USENIX Security Symposium (USENIX Association), 461--478.
\newblock \doarXiv{1901.03597}

\bibitem[{Moore {et~al.}(2021)Moore, Hill, Tildesley, Dyson, \&
  Keeling}]{Moore2021}
Moore, S., Hill, E.~M., Tildesley, M.~J., Dyson, L., \& Keeling, M.~J. 2021,
  The Lancet Infectious Diseases, 21, 793,
  \dodoi{10.1016/S1473-3099(21)00143-2}

\bibitem[{Nakhleh {et~al.}(2021)Nakhleh, Fernandez-Godino, Grosskopf, Wilson,
  Kline, \& Srinivasan}]{Nakhleh2020}
Nakhleh, J.~B., Fernandez-Godino, M.~G., Grosskopf, M.~J., {et~al.} 2021, IEEE
  Transactions on Plasma Science, 49, 2238, \dodoi{10.1109/TPS.2021.3090299}

\bibitem[{Nayak {et~al.}(2020)Nayak, Cowles, Villarini, \& Wafa}]{Nayak2020}
Nayak, M.~A., Cowles, M.~K., Villarini, G., \& Wafa, B.~U. 2020, Water
  Resources Research, 56, e2020WR028256, \dodoi{10.1029/2020WR028256}

\bibitem[{Niemeyer {et~al.}(2020)Niemeyer, Dreicer, \& Stein}]{Niemeyer2020}
Niemeyer, I., Dreicer, M., \& Stein, G., eds. 2020, {Nuclear Non-proliferation
  and Arms Control Verification} (Cham: Springer International Publishing),
  \dodoi{10.1007/978-3-030-29537-0}

\bibitem[{Osborne {et~al.}(2012)Osborne, Roberts, Rogers, \&
  Jennings}]{Osborne2012}
Osborne, M.~A., Roberts, S.~J., Rogers, A., \& Jennings, N.~R. 2012, ACM
  Transactions on Sensor Networks, 9, 1, \dodoi{10.1145/2379799.2379800}

\bibitem[{Osthus {et~al.}(2019)Osthus, {Vander Wiel}, Hoffman, \&
  Wysocki}]{Osthus2019}
Osthus, D., {Vander Wiel}, S.~A., Hoffman, N.~M., \& Wysocki, F.~J. 2019,
  SIAM/ASA Journal on Uncertainty Quantification, 7, 604,
  \dodoi{10.1137/17M1158860}

\bibitem[{Penny(1999)}]{Penny1999}
Penny, W. 1999in  (IEE), 826--831, \dodoi{10.1049/cp:19991214}

\bibitem[{Phillips \& Pohl(2021)}]{Phillips2021}
Phillips, P.~J., \& Pohl, G. 2021, The International Journal of Intelligence,
  Security, and Public Affairs, 23, 34, \dodoi{10.1080/23800992.2020.1834311}

\bibitem[{Pike {et~al.}(2014)Pike, Mackenroth, Hill, \& Rose}]{Pike2014}
Pike, O.~J., Mackenroth, F., Hill, E.~G., \& Rose, S.~J. 2014, Nature
  Photonics, 8, 434, \dodoi{10.1038/nphoton.2014.95}

\bibitem[{Press(1996)}]{Press1996}
Press, W.~H. 1996, arxiv.
\newblock \doarXiv{9604126}

\bibitem[{Pritychenko(2017)}]{Pritychenko2017}
Pritychenko, B. 2017, EPJ Web of Conferences, 146, 01006,
  \dodoi{10.1051/epjconf/201714601006}

\bibitem[{Pritychenko \& Mughabghab(2012)}]{Rajput1992}
Pritychenko, B., \& Mughabghab, S. 2012, Nuclear Data Sheets, 113, 3120,
  \dodoi{10.1016/j.nds.2012.11.007}

\bibitem[{Pritychenko {et~al.}(2012)Pritychenko, Mughabghab, Pritychenko, \&
  Mughabghab}]{Pritychenko2012}
Pritychenko, B., Mughabghab, S.~F., Pritychenko, B., \& Mughabghab, S.~F. 2012,
  NDS, 113, 3120, \dodoi{10.1016/J.NDS.2012.11.007}

\bibitem[{Ramana(2021)}]{Ramana2021}
Ramana, M.~V. 2021, {Beyond our imagination: Fukushima and the problem of
  assessing risk - Bulletin of the Atomic Scientists}.
\newblock
  \url{https://thebulletin.org/2011/04/beyond-our-imagination-fukushima-and-the-problem-of-assessing-risk/}

\bibitem[{Randewich {et~al.}(2020)Randewich, Lock, Garbett, \&
  Bethencourt-Smith}]{Randewich2020}
Randewich, A., Lock, R., Garbett, W., \& Bethencourt-Smith, D. 2020,
  Philosophical Transactions of the Royal Society A, 378, 20200012,
  \dodoi{10.1098/RSTA.2020.0012}

\bibitem[{Reece \& Roberts(2010)}]{Reece2010}
Reece, S., \& Roberts, S. 2010, IEEE Transactions on Aerospace and Electronic
  Systems, 46, 207, \dodoi{10.1109/TAES.2010.5417157}

\bibitem[{Reece {et~al.}(2009)Reece, Roberts, Claxton, \&
  Nicholson}]{Reece2009}
Reece, S., Roberts, S., Claxton, C., \& Nicholson, D. 2009, in 2009 12th
  International Conference on Information Fusion, 1695--1703

\bibitem[{Rhodes(1986)}]{Rhodes1986}
Rhodes, R. 1986, {The making of the atomic bomb} (Simon {\&} Schuster), 886

\bibitem[{Rose {et~al.}(2020)Rose, Hatfield, \& Scott}]{Rose2020}
Rose, S., Hatfield, P., \& Scott, R. 2020, Philosophical transactions. Series
  A, Mathematical, physical, and engineering sciences, 378,
  \dodoi{10.1098/rsta.2020.0014}

\bibitem[{Rose \& Hatfield(2021)}]{Rose2021}
Rose, S.~J., \& Hatfield, P.~W. 2021, Contemporary Physics, 1,
  \dodoi{10.1080/00107514.2021.1959097}

\bibitem[{Ruby {et~al.}(2021)Ruby, Gaffney, Rygg, Ping, \& Collins}]{Ruby2021}
Ruby, J.~J., Gaffney, J.~A., Rygg, J.~R., Ping, Y., \& Collins, G.~W. 2021,
  Physics of Plasmas, 28, 032703, \dodoi{10.1063/5.0040616}

\bibitem[{Sangero \& Halpert(2007)}]{Sangero2007}
Sangero, B., \& Halpert, M. 2007, Jurimetrics, 48, 43.
\newblock \url{http://www.clb.ac.il/uploads/sangero4.pdf}

\bibitem[{Shayler(2000)}]{Shayler2000}
Shayler, D. 2000, {Disasters and Accidents in Manned Spaceflight} (Springer)

\bibitem[{Sinervo(2003)}]{Sinervo2003}
Sinervo, P. 2003, in Conference on Statistical Problems in Particle Physics,
  Astrophysics and Cosmology, Stanford, CA, USA.
\newblock \url{https://cds.cern.ch/record/931829}

\bibitem[{Sivia \& Skilling(2006)}]{Sivia2006}
Sivia, D., \& Skilling, J. 2006, {Data analysis: a Bayesian tutorial} (Oxford
  University Press).
\newblock
  \url{https://global.oup.com/academic/product/data-analysis-9780198568322?cc=us{\&}lang=en{\&}}

\bibitem[{Taleb(2007)}]{Taleb2007}
Taleb, N.~N. 2007, {The Black Swan: Second Edition} (Random House), 480.
\newblock
  \url{http://books.google.co.jp/books?id=GSBcQVd3MqYC{\&}printsec=frontcover{\&}dq=taleb+black+swan{\&}hl={\&}cd=1{\&}source=gbs{\_}api}

\bibitem[{Taleb(2012)}]{Taleb2012}
---. 2012, SSRN Electronic Journal, \dodoi{10.2139/ssrn.1850428}

\bibitem[{Taleb(2018)}]{Taleb2018}
---. 2018, Quantitative Finance, 18, 1.
\newblock \doarXiv{1703.06351}

\bibitem[{Taleb(2020)}]{Taleb2020}
---. 2020.
\newblock \doarXiv{2001.10488}

\bibitem[{{\"{U}}lk{\"{u}}men \& Fox(2011)}]{Fox2011}
{\"{U}}lk{\"{u}}men, G., \& Fox, C.~R. 2011, in Perspectives on Thinking,
  Judging, and Decision Making, ed. H.~{Brun, W., Keren, G., Kirkeb{\o}en, G.,
  {\&} Montgomery} No.~1 (Universitetsforlaget), 14.
\newblock
  \url{https://books.google.com/books/about/Perspectives{\_}on{\_}Thinking{\_}Judging{\_}and{\_}Dec.html?id=9NDiygAACAAJ}

\bibitem[{{Von Toussaint}(2011)}]{VonToussaint2011}
{Von Toussaint}, U. 2011, Reviews of Modern Physics, 83, 943,
  \dodoi{10.1103/RevModPhys.83.943}

\bibitem[{Wagenmakers {et~al.}(2022)Wagenmakers, Sarafoglou, \&
  Aczel}]{Wagenmakers2022}
Wagenmakers, E.-J., Sarafoglou, A., \& Aczel, B. 2022, Nature, 605, 423,
  \dodoi{10.1038/d41586-022-01332-8}

\bibitem[{Walker {et~al.}(1998)Walker, Crowther, Wilkinson, Stokes, \&
  Swinerd}]{Walker1998}
Walker, R., Crowther, R., Wilkinson, J., Stokes, P., \& Swinerd, G. 1998, in
  Mission Design {\&} Implementation of Satellite Constellations (Springer,
  Dordrecht), 317--326, \dodoi{10.1007/978-94-011-5088-0_28}

\bibitem[{Walker(2013)}]{Walker2013}
Walker, S.~G. 2013, Journal of Statistical Planning and Inference, 143, 1621,
  \dodoi{10.1016/j.jspi.2013.05.013}

\bibitem[{Yao {et~al.}(2018)Yao, Vehtari, Simpson, \& Gelman}]{Yao2018}
Yao, Y., Vehtari, A., Simpson, D., \& Gelman, A. 2018, Bayesian Analysis, 13,
  917, \dodoi{10.1214/17-BA1091}

\bibitem[{Zholud(2014)}]{Zholud2014}
Zholud, D. 2014, Bernoulli, 20, \dodoi{10.3150/13-BEJ552}

\end{thebibliography}

\end{document}